\documentclass[reqno,tbtags,intlimits,a4paper,oneside,12pt]{amsart}

\usepackage[foot]{amsaddr}

\usepackage[english]{babel}
\usepackage{amssymb,upref,calc,url}
\usepackage[mathscr]{eucal}
\usepackage{graphicx}
\usepackage{amsmath, amsfonts,amscd,amsthm,amssymb,verbatim}
\usepackage[all]{xy}
\usepackage{textcomp}
\usepackage[in]{fullpage}

\usepackage[title]{appendix}

\newtheorem{theorem}{Theorem}[section]
\newtheorem{corollaryt}[theorem]{Corollary}
\newtheorem{lemma}[theorem]{Lemma}
\theoremstyle{definition}
\newtheorem{definition}[theorem]{Definition}
\newtheorem{remark}[theorem]{Remark}
\newtheorem{example}[theorem]{Example}

\newtheorem{innercustomthm}{Theorem}
\newenvironment{customthm}[1]
  {\renewcommand\theinnercustomthm{#1}\innercustomthm}
  {\endinnercustomthm}

\usepackage{color}

\definecolor{shamrockgreen}{rgb}{0.0, 0.62, 0.38}

\usepackage{hyperref}
\hypersetup{
    colorlinks=true,
    linkcolor=shamrockgreen,
    filecolor=magenta,
    citecolor=blue,
    urlcolor=blue,
    pdftitle={The Neumann--Moser dynamical system and the Korteweg--de Vries hierarchy},
    pdfpagemode=FullScreen,
    }

\begin{document}

\frenchspacing

\title{The Neumann--Moser dynamical system\\
 and the Korteweg--de\;Vries hierarchy}

\author{Polina Baron}

\address{University of Chicago, USA (2024)}
\email{pbaron@uchicago.edu}



\maketitle



\begin{abstract}At the focus of the paper are applications of the well-known Moser transformation of the C. Neumann dynamical system.
It yields us a new quadratic integrable dynamical system on $\mathbb{C}^{3n+1}$, which we call the Neumann--Moser dynamical system.

We present an explicit formula of the inverse of the Moser transformation. Consequently, we obtain explicitly an invertible transformation of the Uhlenbeck--Devaney integrals of the Neumann system into the integrals of our system. One of the main results of the paper is the recurrent solutions of the Neumann--Moser system.

We show that every solution of our system solves the Mumford dynamical system, and vice versa. Every solution of the Neumann--Moser system is proven to solve the stationary Korteweg--de Vries hierarchy. As a corollary, we construct explicit solutions of the Neumann--Moser system in hyperelliptic Kleinian functions.\\
\\
------------------------------------------------------------------------------------------------------------------
\end{abstract}
\vspace{1cm}

\tableofcontents
\newpage

\section*{Introduction}
The Neumann system, sometimes called the C. Neumann system, is among the widest-known and best-studied integrable systems in classical mechanics.
It describes the dynamics of a point particle that is constrained to move on the sphere $S^n$ under the influence of a linear force (equivalently, a quadratic potential).
In 1859, Carl Neumann, a student of Jacobi, showed that \cite{Neum-59}, in the ``physical'' case for $n=2$, the equations of motion can be solved using the Jacobi theory of separation of variables.
In 1877, Rosochatius \cite{Ros-77} proved that one can add a potential given by the sum (with nonnegative weights) of the inverses of the squares of the (Cartesian) coordinates without losing the separability property.
However, it was only a century later that the separability result was generalized to the case of arbitrary $n>2$ by Uhlenbeck \cite{Uhlenbeck-75} and Devaney \cite{Devaney-78}.
Their result was put into a much more accessible form by J{\"u}rgen Moser \cite{Moser-80}, who went on to develop a beautiful theory of the Neumann system (see, among other works, \cite{Moser-79} and \cite{Moser-83}).
The Neumann system was been connected to the geodesic flow on ellipsoids by Kn{\"o}rrer \cite{Knorrer-82}. Twenty years later, Dullin, Richter, Veselov, and Waalkens \cite{DRVW-01} described action integrals of the Neumann system, more convenient for numerical computations. They also showed that the actions of the Neumann system satisfy a Picard--Fuchs equation.

The Neumann system is usually considered in the regime of distinct eigenvalues. In the present paper, we follow the same path. However, the case of coinciding eigenvalues

Moser \cite{Moser-80} and Trubowitz 
discovered a deep connection between the Neumann system and the eigenfunctions problem for the periodic Schr{\"o}dinger operator, which made way for connections of the Neumann system to finite-gap potentials and allowed for the use of the finite-gap integration method for the KdV equation. This method is based on fundamental algebraic properties of equations that admit the Lax representation, which were discovered by Novikov \cite{Nov-74}, who developed it for studying periodic and quasi-periodic multi-soliton analogues of solutions of the Korteweg--de Vries equation. This method was developed further by Dubrovin, Matveev, and Novikov \cite{DMN-76}. Using this method, Veselov \cite{Ves-80} proved the existence of a purely algebraic isomorphism, called the Moser--Trubowitz isomorphism, between the $(n+1)$-dimensional sheaf (universal bundle) of phase spaces that correspond to stationary higher-order Korteweg--de Vries equation and the sheaf (universal bundle) of the Neumann systems with all possible quadratic finite-gap potentials.

Dubrovin and Novikov \cite{DN-74} demonstrated that the universal bundle of Jacobians $J(V)$ of hyperelliptic curves $V$ of genus $g$ (the DN space) with a chosen branching point $E_{2g+2}=\infty$ is birationally equivalent to $\mathbb{C}^{3g+1}$.  Therefore, one obtains a polynomial integrable dynamical system on $\mathbb{C}^{3g+1}$ with $g$ commuting flows along the images of the Jacobians of hyperelliptic curves that are distinguished by the values of the integrals of the finite-gap Korteweg–-de Vries hierarchy.

Mumford \cite{Mumford-84} developed the theta-functional theory of the universal bundle of Jacobians $J(V)$. As one of the core tools, he introduced a new dynamical system, which was later named after him (see \cite[\S 3]{Mumford-84}, also \cite[Section 1]{Buch23}). He described the Moser transformation of the Neumann system as one of its applications (see \cite[p. 3.57]{Mumford-84}). Detailed exploration and generalizations of the Mumford system are given by Vanhaecke in \cite[Chapter VI]{Vanhaecke-96}. Buchstaber \cite{Buch23} developed the differential-algebraic theory of the Mumford system and described its solutions in hyperelliptic Kleinian functions (see also \cite{Bar-23}).

\vspace{.25cm}
Our paper investigates the Moser transformation of the complexified version of the Neumann dynamical system whose $n+1$ parameters are distinct. This transformation is known to be a degree $2^{n+1}$ branched covering over $\mathbb{C}^{3n+1}$ (see, for example, \cite[p. 3.59, Lemma 4.3]{Mumford-84}). Its application leads us to a new integrable dynamical system on $\mathbb{C}^{3n+1}$, which we call the Neumann--Moser dynamical system (see Definition \ref{NMdef} and Theorem \ref{th1}). This system is quadratic and has a representation in the Lax form (see Theorem \ref{th2}). It possesses $2n+1$ canonical integrals $h_1,\,\ldots,\,h_{2n+1}$.

We present to the reader 5 core results. All of them are given and proved by explicit and complete formulas.

\begin{customthm}{1}[See Lemmas \ref{Mos-lemma-1} and \ref{Mos-lemma-2}, Corollary \ref{Moser-cor}]
The Moser transformation is invertible up to the sheet of the branched covering. As a corollary, solutions of the Neumann system can be obtained from solutions of the Neumann--Moser system.
\end{customthm}


\begin{customthm}{2}[See Theorem \ref{F-h_th}]
The integrals $h_{n+1},\,\ldots,\,h_{2n+1}$ are linear combinations of the integrals $h_{1},\,\ldots,\,h_{n}$ with coefficients that are polynomials in the parameters $a_1,\,\ldots,\,a_{n+1}$ of the corresponding Neumann system.\\
There exists an invertible transformation of the integrals $h_1,\,\ldots,\,h_{n}$ to the Uhlenbeck--Devaney integrals $F_1,\,\ldots,\,F_{n+1}$ of the Neumann system ($\sum_{k=1}^{n+1}F_k=1$). This transformation is linear in $h_1,\,\ldots,\,h_{n}$ and depends polynomially on $a_1,\,\ldots,\,a_{n+1}$.
\end{customthm}

\begin{customthm}{3}[See Theorem \ref{u-rec-th} and Corollary \ref{u-rec-cor-2}]
There exists a recursion of differential polynomials that reduces the problem of constructing solutions to the Neumann--Moser system to the Cauchy problem for an ordinary differential equation of order $2n$ that is dependent on $n+1$ parameters $h_1,\,\ldots,\,h_{n+1}$.
\end{customthm}

\begin{customthm}{4}[See Theorem \ref{M-NS-th}]
Every solution of the Neumann--Moser system can be extended to a solution of the Mumford dynamical system, and every solution of the Mumford dynamical system solves the Neumann--Moser system in $\partial_1=\partial=\frac{\partial}{\partial t}$.
\end{customthm}

\begin{customthm}{5}[See Theorems \ref{KdV-NS-th} and \ref{KdV-hkf-th}]
Every solution of the Neumann--Moser system solves the $(n+1)$-stationary Korteweg--de Vries hierarchy. Consequentially, under certain conditions, the Neumann-Moser system can be solved in hyperelliptic Kleinian functions.
\end{customthm}

The paper is organized as follows.\\
Section\;\ref{sect1} gives the formal description of the C. Neumann dynamical system.\\
Section\;\ref{sect2} constructs the Moser transformation as a degree $2^{n+1}$ branched covering and provides explicit formulas of its inverse (up to a sheet).\\
Section\;\ref{sect3} introduces the Neumann--Moser dynamical system and conclude the proof of its birational equivalence to the Neumann dynamical system up to a sheet.\\
Section\;\ref{sect4} explicitly constructs and relates to each other the integrals of the Neumann and the Neumann--Moser systems.\\
Section\;\ref{sect5} obtains the recurrent solutions of the Neumann--Moser system.\\
Section\;\ref{sect6} proves that every solution of the Mumford dynamical system solves the Neumann--Moser dynamical system in the first time variable, and vice versa. \\
Section\;\ref{sect7} relates the solution of the Neumann--Moser dynamical system to solution of the $(n+1)$-stationary Korteweg--de Vries hierarchy and obtains explicit solutions of the Neumann--Moser system in hyperelliptic Kleinian functions as a consequence.\\
Appendix \ref{sectA} provides necessary results on matrices whose entries are symmetric polynomials, which frequently appear in the formulas in this paper.

\vspace{.5cm}

\textbf{Acknowledgements:} The author is deeply grateful to S.~Filip for supporting the author's breadth of mathematical interests and inadvertently sparkling the inquiry that led to this paper. The author thanks to V.~M.~Buchstaber for his unwavering encouragement, meaningful advice, and numerous discussions. The author is grateful to M.~Adler and V.~N.~Roubtsov for stimulating conversations.

\section{The Neumann dynamical system}\label{sect1}
In 1859, C.~Neumann \cite{Neum-59} introduced
\begin{definition}
The ($n$-dimensional) \emph{Neumann dynamical system} describes the movement of a particle on a real unitary $n$-sphere
under the influence of a potential that is a quadratic form. It can be described by the system of ordinary differential equations
\begin{equation}\label{N-sys}
\begin{cases}
  \dot{x}_k(t)=y_k(t) \\
  \dot{y}_k(t)=-\big(\Gamma(t)+a_k\big)x_k(t),
\end{cases} \, 1\leq k \leq n+1,
\end{equation}
with the vector of parameters $A=(a_1,\,\ldots,\,a_{n+1})\in\mathbb{R}^{n+1}$ such that $a_i\neq a_j$ if $i\neq j$,
the correction $\Gamma(t)=\sum\limits^{n+1}_{i=1}(y_i^2-a_ix_i^2)$, and with initial conditions
\[
\sum^{n+1}_{i=1}x_i^2(0) = 1, \quad \sum^{n+1}_{i=1}x_i(0)y_i(0) = 0.
\]
\end{definition}
We will further drop writing the dependence on $t$ where it is obvious for the sake of brevity.

The correction by $-\Gamma$ in the system \eqref{N-sys} models the force perpendicular to the sphere that keeps the particle on this sphere.
In other words, it causes every solution $(x(t),y(t))$ of the system \eqref{N-sys} with the given initial conditions to satisfy
\[
\sum\limits^{n+1}_{k=1}x_k^2(t) \equiv 1\; \text{ and }\; \sum\limits^{n+1}_{k=1}x_k(t)y_k(t) \equiv 0,
\]
%
It is worth noting that the value of $\Gamma$ arises naturally in the real case. For the function $\varphi = 1-\sum\limits^{n+1}_{k=1}x_k^2$,
direct calculations show that $\dot{\varphi}=-\sum\limits^{n+1}_{k=1}x_ky_k$ and
that the function $\varphi$ satisfies the differential equation $\ddot{\varphi} = -2\Gamma \varphi$ with initial conditions
$\varphi(0)=0$ and $\dot{\varphi}(0)=0$. Therefore, $\varphi(t)\equiv 0$.

\vspace{.5cm}

The Neumann dynamical system can be generalized to the case of complex coordinates. Consider $\mathbb{C}^{2n+2}=\mathbb{C}^{n+1}\times\mathbb{C}^{n+1}$
with coordinates $X=(x_1,\ldots,x_{n+1})$ and $Y=(y_1,\ldots,y_{n+1})$
as a symplectic manifold with a canonical Poisson bracket $\{ \cdot,\cdot \}$ for which $\{ x_i,y_i \}=1$ and all other values
of the coordinate brackets are zero. Let us introduce the algebraic $2n$-dimensional manifold
\[
\mathcal{M}_x = \Big\{ (X,Y)\in \mathbb{C}^{2n+2}\;:\; \sum\limits^{n+1}_{k=1}x_k^2 = 1, \; \sum\limits^{n+1}_{k=1}x_ky_k = 0 \Big\}
\]
with a Poisson bracket induced by the bracket in $\mathbb{C}^{2n+2}$.

The calculations related to the function $\varphi$ defined above remain unchanged, meaning that the Neumann system in complex coordinates is well-defined,
and all of its solutions lie on the manifold $\mathcal{M}_x$.
It is well-known that system \eqref{N-sys} defined on $\mathcal{M}_x$ is a Hamiltonian system with the Hamiltonian
$$
H(X,Y) = \frac{1}{2}\sum\limits^{n+1}_{k=1}(a_kx_k^2+y_k^2).
$$
Vectors of parameters $A=(a_1,\ldots,a_{n+1})$ form the manifold
\[
\mathcal{A} = \Big\{ A\in \mathbb{C}^{n+1}\;:\; a_i\neq a_j \text{ if }i\neq j \Big\}.
\]
Uhlenbeck \cite{Uhlenbeck-75} and Devaney \cite{Devaney-78} showed that the Hamiltonian $H(X,Y)$ has $n+1$ algebraic integrals in the involution (i.e. Poisson commuting)
\[
F_k(X,Y) = x_k^2 + \sum_{l\neq k}\frac{(x_ky_l-x_ly_k)^2}{a_k-a_l},\quad \text{ where }\; \sum\limits^{n+1}_{k=1}F_k(X,Y)=1.
\]
Thus, this Hamiltonian system is completely Liouville integrable.

\section{The Moser transformation}\label{sect2}
\begin{definition}\label{mtdef}
The \emph{Moser transformation}
$$
\mathbf{M}_a\,:\,(X,Y)\to (\mathbf{U},\mathbf{V},\mathbf{W}),
$$
where the vectors $\mathbf{U}=(\mathbf{u}_1,\ldots,\mathbf{u}_{n})$, $\mathbf{V}=(\mathbf{v}_1,\ldots,\mathbf{v}_{n})$, $\mathbf{W}=(\mathbf{w}_1,\ldots,\mathbf{w}_{n+1})$
are defined by the coefficients of the generating polynomials
\begin{align}\label{moser-transform}
\mathbf{U}_\xi &=\; \xi^n\, +\, \sum_{i=1}^{n}\mathbf{u}_i\xi^{n-i} \; =\; f_\xi\sum_{j=1}^{n+1}\frac{x_j^2}{\xi-a_j},\\
\mathbf{V}_\xi &=\; \sum_{i=1}^{n}\mathbf{v}_i\xi^{n-i} \quad =\; \sqrt{-1}f_\xi\sum_{j=1}^{n+1}\frac{x_jy_j}{\xi-a_j},\\
\mathbf{W}_\xi &=\; \xi^{n+1} + \sum_{i=1}^{n+1}\mathbf{w}_i\xi^{n+1-i} = f_\xi\sum_{j=1}^{n+1}\Big(\frac{y_j^2}{\xi-a_j}+1\Big),
\end{align}
where $f_\xi =\; \prod_{j=1}^{n+1}(\xi-a_j).$
\end{definition}
\begin{remark}
David Mumford \cite[p. 3.57]{Mumford-84} attributed this transformation to Moser, but never cited any paper of his. The authors of the present paper did not succeed in finding this transformation in Moser's works.
\end{remark}

For the sake of convenience, we will consider the generalized Moser transformation $\mathbf{M}:E_x\to E_u$,
where $E_x=\mathcal{A}\times\mathcal{M}_x$ is the trivial bundle with an $n+1$-dimensional base $\mathcal{A}$
and a $2n$-dimensional fiber $\mathcal{M}_x$ immersed into $\mathbb{C}^{2n+2}$, and $E_u=\mathcal{A}\times\mathcal{M}_u$
is a trivial bundle with an $n+1$-dimensional base $\mathcal{A}$ and a (yet unknown to us) fiber $\mathcal{M}_u$ immersed into $\mathbb{C}^{3n+1}$.
The transformation $\mathbf{M}$ acts trivially on the base $\mathcal{A}$.

Let us rewrite the Moser transformation in a more explicit form.

First, note that the result of the Moser transformation is not changed by the involutions $i_k:(X,Y)\to (X,Y)$, $1\leq k\leq n+1$,
that multiply $x_k$ and $y_k$ by $-1$ and leave all other coordinates unchanged.
Therefore, the Moser transformation can be decomposed into two transformations, the first of which is a degree $2^{n+1}$ branched covering $\pi_{Inv}: E_x\to \widetilde{E}_x$,
where $\widetilde{E}_x=\mathcal{A}\times\widetilde{\mathcal{M}}_x$ is the result of factoring the bundle $E_x$ by the action of the group $Inv=\langle i_1,\,\ldots,\,i_{n+1} \rangle$.
We can write the map $\pi_{Inv}$ in the following form: $\pi_{Inv}: (X,Y)\to (XX, XY, YY)$, where $XX=X\odot X$, $XY=Y\odot Y$, $YY=Y\odot Y$
are Hadamard products of $X$ and $Y$. Here, the fiber $\widetilde{\mathcal{M}}_x$ is considered to be a $2n$-dimensional manifold immersed into $\mathbb{C}^{3n+3}$.

For the sake of convenience, we introduce the embedding $\mathrm{Im}_1: \widetilde{\mathcal{M}}_x\hookrightarrow \mathbb{C}^{3n+6}$
defined by the addition of an extra constant coordinate to each of the vectors $XX,\,XY,\,YY$ in the following way:
\begin{align}
  \mathrm{Im}_1(XX) &=\; (0,\,x_1^2,\,\ldots,\,x_{n+1}^2), \\[2pt]
  \mathrm{Im}_1(XY) &=\; (0,\,x_1y_1,\,\ldots,\,x_{n+1}y_{n+1}), \\[2pt]
  \mathrm{Im}_1(YY) &=\; (1,\,y_1^2,\,\ldots,\,y_{n+1}^2).
\end{align}

Additionally, let us introduce the embedding $\widetilde{\mathcal{M}}_u\hookrightarrow \mathbb{C}^{3n+6}$ defined by addition of $2$ extra constant coordinates
to each of the vectors $\mathbf{U},\,\mathbf{V},\,\mathbf{W}$ in the following way:
\begin{align}
  \mathrm{Im}_2(\mathbf{U}) &=\; (0,\,1,\,\mathbf{u}_1,\,\ldots,\,\mathbf{u}_n), \\[2pt]
  \mathrm{Im}_2(\mathbf{V}) &=\; (0,\,0,\,\mathbf{v}_1,\,\ldots,\,\mathbf{v}_n), \\[2pt]
  \mathrm{Im}_2(\mathbf{W}) &=\; (1,\,\mathbf{w}_1,\,\mathbf{w}_2,\,\ldots,\,\mathbf{w}_{n+1}).
\end{align}

Direct calculations show that, for a fixed vector of parameters $A$,
\begin{equation}\label{f-13}
\mathrm{Im}_2(\mathbf{U}) = M_A\cdot \mathrm{Im}_1(XX), \quad
\mathrm{Im}_2(\mathbf{V}) = \sqrt{-1}M_A\cdot \mathrm{Im}_1(XY), \quad
\mathrm{Im}_2(\mathbf{W}) = M_A\cdot \mathrm{Im}_1(YY),
\end{equation}
where $M(-A)$ is a $(n+2)\times(n+2$)-matrix
\begin{equation}\label{M_A}
M_A=M(-A)=
     \left(\begin{array}{ccccc}
    1 & 0 & 0 & \ldots & 0 \\
    e_1 & 1 & 1 & \ldots & 1 \\
    e_2 & e_1^{(1)} & e_1^{(2)} & \cdots & e_1^{(n+1)} \\
    \vdots & \vdots & \vdots & \ddots & \vdots \\
    e_{n+1} & e_n^{(1)} & e_n^{(2)} & \cdots & e_n^{(n+1)}
    \end{array}\right).
\end{equation}
Here, $e_k=e_k(-A)$,\; $e_k^{(l)}=e_k^{(l)}(-A)$ stand for the elementary symmetric polynomials in variables $-a_1,\,\ldots,\,-a_n$. For the formal definition, please refer to Definition \ref{symmetricdef1} in Appendix A.
\begin{example}For the cases $n=1$ and $n=2$, we have
\begin{align*}
M_1&\,=\,M(-a_1,\,-a_2)\,=\,
\begin{pmatrix}
    1 & 0 & 0 \\
    -a_1-a_2 & 1 & 1 \\
    a_1a_2 & -a_2 & -a_1
    \end{pmatrix},
    \quad
    \text{and}
    \quad\\
M_2&\,=\,M(-a_1,\,-a_2,\,-a_3)\,=\,
     \begin{pmatrix}
    1 & 0 & 0 &0 \\
    -a_1-a_2-a_3 & 1 & 1 & 1\\
    a_1a_2+a_2a_3+a_3a_1 & -a_2-a_3 & -a_1-a_3 & -a_2-a_1\\
    -a_1a_2a_3 & a_2a_3 & a_1a_3 & a_3a_2
    \end{pmatrix}.
\end{align*}
Note that $\det M_1\,=\,a_2 - a_1$ and $\det M_2\,=\,(a_2-a_1)(a_3-a_1)(a_3-a_2)$.
\end{example}
Summing up, the generalized Moser transformation $\mathbf{M}:E_x\to E_u$ is a degree $2^{n+1}$ branched covering that can be rewritten in the form
$$\mathbf{M}(A;X,Y)=\mathrm{Im}_2^{-1}\circ M(-A)\circ \mathrm{Im}_1\circ\pi_{Inv}.$$

\begin{example}Computing explicitly, we get
\begin{multline} \label{u_1}
\;\; \mathbf{u}_1 =\,(e_2,\,e_1^{(1)},\,e_1^{(2)},\,\ldots,\,e_1^{(n+1)})\times (0,x_1^2,\,\ldots,\,x_{n+1}^2)^T=\sum_{k=1}^{n+1}e_1^{(k)}x_k^2= \\
=\sum_{k=1}^{n+1}x_k^2\Big(\sum_{i=1}^{n+1}(-a_i)-(-a_k)\Big)=-\sum_{i=1}^{n+1}a_i\sum_{k=1}^{n+1}x_k^2+\sum_{k=1}^{n+1}a_kx_k^2
=\sum_{k=1}^{n+1}a_k(x_k^2-1),
\end{multline}
\begin{align}
\mathbf{v}_1 &=\; \sqrt{-1}(e_2,e_1^{(1)},\ldots,e_1^{(n+1)})\times (0,x_1y_1,\ldots,x_{n+1}y_{n+1})^T
= -\sqrt{-1}\sum_{k=1}^{n+1}a_kx_ky_k, \label{v_1}\\
\mathbf{w}_1 &=\; (e_1,\,1,\,\ldots,\,1)\times (1,\,y_1^2,\,\ldots,\,y_{n+1}^2)^T
=e_1+\sum_{k=1}^{n+1}y_k^2 = \sum_{k=1}^{n+1}(y_k^2-a_k). \label{w_1}
\end{align}
\end{example}

\begin{lemma}\label{glemma}
\begin{equation}\label{g-formula}
\Gamma =\mathbf{w}_1-\mathbf{u}_1.
\end{equation}
\end{lemma}
\begin{proof}From Equations \eqref{u_1} and \eqref{w_1},
\begin{equation*}
\Gamma = \sum\limits^{n+1}_{i=1}y_i^2 - \sum\limits^{n+1}_{i=1}a_ix_i^2
=\Big(\mathbf{w}_1+\sum^{n+1}_{i=1}a_i\Big)-\Big(\mathbf{u}_1+\sum^{n+1}_{i=1}a_i\Big)= \mathbf{w}_1-\mathbf{u}_1.
\end{equation*}
\vspace{-.1cm}
\end{proof}

\begin{lemma}\label{Mos-lemma-1}
  The matrix $M(-A)$ is invertible for all $A\in\mathcal{A}$. Its inverse is
$$
M^{-1}_A=\frac{1}{\Pi_A}
\left(\begin{array}{cccc}
\Pi_A & \ldots & 0 & 0 \\
-a_1^{n+1}\Pi^{(1)}_A & \ldots & -a_1\Pi^{(1)}_A & -\Pi^{(1)}_A\\
a_2^{n+1}\Pi^{(2)}_A &  \ldots & a_2\Pi^{(2)}_A & \Pi^{(2)}_A \\
\vdots &  \ddots & \vdots & \vdots \\
(-1)^{n+1}a_{n+1}^{n+1}\Pi^{(n+1)}_A & \ldots & (-1)^{n+1}a_{n+1}\Pi^{(n+1)}_A & (-1)^{n+1}\Pi^{(n+1)}_A
\end{array}\right),
$$
where $\Pi_A=\Pi(-A)$,\; $\Pi^{(l)}_A=\Pi^{(l)}(-A)$ are Vandermonde polynomials from Definition \ref{symmetricdef2}.
\end{lemma}
The fact that $M^{-1}_A$ has this form is proven in Lemma \ref{inverse_M_lemma}.
\begin{example}Denote $\det M_n$ by $\mathrm{D}_n$, $n\geq 1$. For the cases $n=1$ and $n=2$, we have
\begin{align*}
    M^{-1}_1&\,=\,
    \frac{1}{\mathrm{D}_1}\begin{pmatrix}
    D_1 & 0 & 0 \\
    -a_1^2 & -a_1 & -1 \\
    a_2^2 & a_2 & 1
    \end{pmatrix},
    \quad
    \text{and}
    \quad\\
    M^{-1}_2&\,=\,
    \frac{1}{\mathrm{D}_2}\begin{pmatrix}
    D_2 & 0 & 0 & 0\\
    -a_1^3(a_3-a_2) & -a_1^2(a_3-a_2) & -a_1(a_3-a_2) & -(a_3-a_2) \\
    a_2^3(a_3-a_1) & a_2^2(a_3-a_1) & a_2(a_3-a_1) & a_3-a_1\\
    -a_3^3(a_2-a_1) & -a_3^2(a_2-a_1) & -a_3(a_2-a_1) & -(a_2-a_1)
    \end{pmatrix}
    .\end{align*}
\end{example}
\vspace{.15cm}

From \eqref{f-13}, one obtains
\begin{lemma}\label{Mos-lemma-2}
\begin{equation*}\label{f-13-1}
\mathrm{Im}_1(XX) = M_A^{-1}\cdot \mathrm{Im}_2(\mathbf{U}), \;
\mathrm{Im}_1(XY) = -\sqrt{-1}\;M_A^{-1}\cdot \mathrm{Im}_2(\mathbf{V}), \;
\mathrm{Im}_1(YY) = M_A^{-1}\cdot \mathrm{Im}_2(\mathbf{W}).
\end{equation*}
\end{lemma}
Rewriting these vector equations in the form of regular equations, we get
\begin{corollaryt}\label{Moser-cor}
\begin{align}
x_i^2
  &=\;(-1)^{i-1}\Pi^{(i)}_Aa_i^{n} +(-1)^{i-1}\Pi^{(i)}_A \sum_{k=1}^{n}a_i^{n-k}\mathbf{u}_{k},\\
x_iy_i &=\;-\sqrt{-1}(-1)^{i-1}\Pi^{(i)}_A \sum_{k=1}^{n}a_i^{n-k}\mathbf{v}_{k},\\
y_i^2
  &=\;(-1)^{i-1}\Pi^{(i)}_Aa_i^{n+1} +(-1)^{i-1}\Pi^{(i)}_A \sum_{k=1}^{n}a_i^{n+1-k}\mathbf{w}_{k}.
\end{align}
\end{corollaryt}

\section{The Neumann--Moser dynamical system}\label{sect3}\text{}\\
The Moser transformation $\mathbf{M}_A$ of the Neumann system give rise to a new system of ordinary differential equations. In this section, we describe this new system explicitly. To begin with, we would like to introduce a new
\begin{definition}\label{NMdef}
The \emph{Neumann--Moser dynamical system} is a dynamical system in $\mathbb{C}^{3n+1}$ with coordinates $U(t)=(u_1,\ldots,u_{n})$, $V(t)=(v_1,\ldots,v_{n})$,
$W(t)=(w_1,\ldots,w_{n+1})$ defined by the system of differential equations
\begin{align}
\dot{U}_\xi &=\; -2V_\xi, \label{f-23} \\[2pt]
\dot{V}_\xi &=\; -(\xi+w_1-u_1)U_\xi + W_\xi, \label{f-24} \\[2pt]
\dot{W}_\xi &=\; \; 2(\xi+w_1-u_1)V_\xi \label{f-25}
\end{align}
on generating polynomials 
\begin{equation*}
U_\xi =\; \xi^n\, +\, \sum_{i=1}^{n}u_i\xi^{n-i},\quad
V_\xi =\; \sum_{i=1}^{n}v_i\xi^{n-i},\quad
W_\xi =\; \xi^{n+1} + \sum_{i=1}^{n+1}w_i\xi^{n+1-i}.
\end{equation*}
\end{definition}

\begin{theorem}\label{th1}
Let $(\mathbf{U},\,\mathbf{V},\,\mathbf{W})$ be the result of applying the Moser transformation to the Neumann system.
Let $\tau=\sqrt{-1}t$. Define new vectors $U,\,V,\,W$ such that $U(t)=\mathbf{U}(\tau)$, $V(t)=\mathbf{V}(\tau)$, $W(t)=\mathbf{W}(\tau)$. Then
$U(t),\,V(t),\,W(t)$ give the Neumann--Moser system (\eqref{f-23}--\eqref{f-25}).
\end{theorem}
\begin{remark}
The switch from the real time to the imaginary one is, admittedly, confusing. We were likewise surprised to find out that its choice considerably simplifies the explicit forms of relations between the Neumann--Moser system and other systems considered in the following sections of the paper. The reason for it is the opportunity to get rid of the coefficient $\sqrt{-1}$ and get a system of differential equations with purely real constants. Indeed, for any function $F$ dependent on $t$ one has
$$\dot{F}(\tau)=\frac{\partial}{\partial t} F(\tau)=
\frac{\partial F(\tau)}{\partial \tau}\;  \frac{\partial \tau}{\partial t}=
\sqrt{-1}\frac{\partial}{\partial t} F(t)=\sqrt{-1}\dot{F}(t),$$
\end{remark}
\begin{proof}
Direct computations show that
  \begin{align*}
\dot{U}_\xi &=\dot{\mathbf{U}}_\xi(\sqrt{-1}t) =\, -\sqrt{-1}f_\xi \sum_{j=1}^{n+1}\frac{2x_j\dot{x}_j}{\xi-a_j} = -2\sqrt{-1}f_\xi\sum_{j=1}^{n+1}\frac{x_jy_j}{\xi-a_j} = -2V_\xi,\\
\dot{V}_\xi &=\dot{\mathbf{V}}_\xi(\sqrt{-1}t) =\, -\sqrt{-1}\sqrt{-1}f_\xi\sum_{j=1}^{n+1}\frac{x_j\dot{y}_j+\dot{x}_jy_j}{\xi-a_j} =\\ &=f_\xi\sum_{j=1}^{n+1}\frac{-x_j^2(\Gamma +a_i)+y_j^2}{\xi-a_j} =
 -\Gamma U_\xi+W_\xi-f_\xi\Big(\sum_{j=1}^{n+1}\frac{a_jx_j^2}{\xi-a_j}+1\Big),\\
\dot{W}_\xi &=\dot{\mathbf{W}}_\xi(\sqrt{-1}t) =\, -\sqrt{-1}f_\xi\sum_{j=1}^{n+1}\frac{2y_j\dot{y}_j}{\xi-a_j} = \\
&= -2\sqrt{-1}f_\xi\sum_{j=1}^{n+1}\frac{-x_jy_j(\Gamma +a_j)}{\xi-a_j} =
2\Gamma V_\xi+2\sqrt{-1}f_\xi\sum_{j=1}^{n+1}\frac{a_jx_jy_j}{\xi-a_j}.
\end{align*}
\vspace{.25cm}

\begin{lemma}
$$
\dot{V}_\xi = -(\Gamma +\xi)U_\xi + W_\xi.
$$
\end{lemma}

\begin{proof}
For each $1\leq k\leq n-1$, the coefficient of $\xi^{n-k}$ in $f_\xi+f_\xi\sum\limits_{j=1}^{n+1}\frac{a_jx_j^2}{\xi-a_j}$\; is
$$
e_{k+1}(\!-A)+\sum_{j=1}^{n+1}a_jx_j^2e_k^j(\!-A)\!=e_{k+1}(A)+\sum_{j=1}^{n+1}x_j^2\Big(\!-e_{k+1}(\!-A)+e_{k+1}^j(\!-A)\Big)\!=\!\sum_{j=1}^{n+1}x_j^2e_{k+1}^j(\!-A).
$$
The coefficient of $\xi^{n+1}$ is $1$. The coefficient of $\xi^{n}$ is $\sum\limits_{j=1}^{n+1}a_j(x_j^2-1).$
The coefficient of $\xi^{0}$ is
$$
\sum_{j=1}^{n+1}a_jx_j^2(-1)^n\prod_{i=1}^{n+1}\Big(a_i\frac{1}{a_j}\Big)+(-1)^{n+1}\prod_{i=1}^{n+1}a_i=0.
$$
Then
$$
f_\xi\Big(\sum_{j=1}^{n+1}\frac{a_jx_j^2}{\xi-a_j}+1\Big)=\xi f_\xi\sum_{j=1}^{n+1}\frac{x_j^2}{\xi-a_j}=\xi U_\xi.
$$
\end{proof}

\begin{lemma}
$$
\dot{W}_\xi = 2(\Gamma +\xi)V_\xi.
$$
\end{lemma}
\begin{proof}
For each $1\leq k\leq n-1$, the coefficient of $\xi^{n-k}$ in $2\sqrt{-1}f_\xi\sum\limits_{j=1}^{n+1}\frac{a_jx_jy_j}{\xi-a_j}$\; is
\begin{multline*}
2\sqrt{-1}\sum_{j=1}^{n+1}a_jx_jy_je_k^j(-A)=2\sqrt{-1}\sum_{j=1}^{n+1}a_jx_jy_j\Big(-e_{k+1}(-A)+e_{k+1}^j(-A)\Big)= \\
=2\sqrt{-1}\sum_{j=1}^{n+1}x_jy_je_{k+1}^j(-A).
\end{multline*}
The coefficient of $\xi^{0}$ is
$$
2\sqrt{-1}\sum_{j=1}^{n+1}a_jx_jy_j(-1)^n\prod_{i=1}^{n+1}\Big(a_i\frac{1}{a_j}\Big)=0.
$$
Then
$$
2\sqrt{-1}f_\xi\sum_{j=1}^{n+1}\frac{a_jx_jy_j}{\xi-a_j}=2\xi V_\xi.
$$
\end{proof}

Applying Lemma \ref{glemma} to $\Gamma$, we arrive at the Neumann--Moser system \eqref{f-23}--\eqref{f-25}, concluding the proof of Theorem \ref{th1}.
\end{proof}

For the sake of convenience, we would like to present Equations \eqref{f-23}--\eqref{f-25} governing the Neumann--Moser system in coordinate form.

\begin{lemma}
  \begin{align}
\dot{u}_i&=-2v_i\quad\forall\; 1\leq i\leq n; \label{u-i}\\[2pt]
\dot{v}_i&=(u_1-w_1)u_{i}-u_{i+1}+w_{i+1}=-\Gamma u_i-u_{i+1}+w_{i+1}\quad\forall\; 1\leq i\leq n-1, \label{v-i}\\[2pt]
\dot{v}_n&=(u_1-w_1)u_n+w_{n+1}=-\Gamma u_n+w_{n+1}; \label{v-n}\\[2pt]
\dot{w}_1&=2v_1, \label{w-1}\\[2pt]
\dot{w}_i&=2v_i+2(w_1-u_1)v_{i-1}=2v_i+2\Gamma v_{i-1}\quad\forall\; 2\leq i\leq n-1, \label{w-i}\\[2pt]
\dot{w}_{n+1}&=2(w_1-u_1)v_n=2\Gamma u_n. \label{w-n}
\end{align}
\end{lemma}
\begin{proof}
\[
\sum_{i=1}^{n}\dot{u}_i\xi^{n-i} = \dot{U}_\xi=-2\dot{V_xi}=-2\sum_{i=1}^{n}v_i.
\]
\begin{multline*}
\sum_{i=1}^{n}\dot{v}_i\xi^{n-i}=\dot{V}_\xi=-(\xi+w_1-u_1)U_\xi+W_\xi= \\
=-(\xi+w_1-u_1)\Big(\xi^n\, +\, \sum_{i=1}^{n}u_i\xi^{n-i}\Big)+\xi^{n+1} + \sum_{i=1}^{n+1}w_i\xi^{n+1-i}=\\
=-\xi^{n+1}- \sum_{i=1}^{n}u_i\xi^{n+1-i}+(u_1-w_1)\xi^n + (u_1-w_1)\sum_{i=1}^{n}u_i\xi^{n-i}+\xi^{n+1}+\sum_{i=1}^{n+1}w_i\xi^{n+1-i}=\\
=(u_1-w_1)\xi^n + (u_1-w_1)\sum_{i=1}^{n}u_i\xi^{n-i}  + \sum_{i=0}^{n-1}(w_{i+1}-u_{i+1})\xi^{n-i}+w_{n+1}=\\
=(u_1-w_1)\xi^n +(w_1-u_1)\xi^{n}+ \sum_{i=1}^{n-1}(u_{i}(u_1-w_1)-u_{i+1}+w_{i+1})\xi^{n-i} + (u_1-w_1)u_{n}+w_{n+1}.
\end{multline*}
\begin{multline*}
\sum_{i=1}^{n+1}\dot{w}_i\xi^{n+1-i} = \dot{W}_\xi=2(\xi+w_1-u_1)V_\xi=2(\xi+w_1-u_1)\sum_{i=1}^{n}v_i\xi^{n-i}=\\
=2\sum_{i=1}^{n}v_i\xi^{n+1-i}+2(w_1-u_1)\sum_{i=1}^{n}v_i\xi^{n-i} = 2\sum_{i=1}^{n}v_i\xi^{n+1-i}+2(w_1-u_1)\sum_{i=2}^{n+1}v_{i-1}\xi^{n+1-i}=\\
=2v_1\xi^{n}+\sum_{i=1}^{n}2(v_i+w_1v_{i-1}-u_1v_{i-1})\xi^{n+1-i}+2(w_1-u_1)v_n.
\end{multline*}
\end{proof}
\begin{corollaryt}\label{vw-u_cor}
All coordinates $v_i$ and $w_i$ can be written in the form of differential polynomials in $\Gamma,\,u_1,\,\ldots,\,u_n$. Explicitly,
  \begin{align}
    v_i  &= -\frac{1}{2}\dot{u}_i \;\;\forall\, 1\leq i\leq n;\\
     w_1 &= \Gamma+u_1,\quad
    w_{i}= -\frac{1}{2}\ddot{u}_{i-1}+\Gamma u_{i-1}+u_{i}\;\; \forall\, 2\leq i\leq n,\quad
    w_{n+1}=-\frac{1}{2}\ddot{u}_{n}+\Gamma u_{n}.
  \end{align}
\end{corollaryt}

The Neumann--Moser system in terms of generating polynomials has a very convenient representation in the Lax form.
Let us put
\begin{equation}\label{laxeq}
L_\xi =
\begin{pmatrix}
  V_\xi & U_\xi \\
  W_\xi & -V_\xi
\end{pmatrix}, \qquad
K_\xi =
\begin{pmatrix}
  0 & -1 \\
  -(\xi+w_1-u_1) & 0
\end{pmatrix}.
\end{equation}
\begin{theorem}\label{th2}The pair of matrices $L_\xi$ and $K_\xi$ satisfy Lax's equation
\[
\dot{L}_\xi = [L_\xi,K_\xi].
\]
\end{theorem}

\begin{proof}
\begin{multline*}
[L_\xi,K_\xi]=
\begin{pmatrix}
  V_\xi & U_\xi \\
  W_\xi & -V_\xi
\end{pmatrix}
\begin{pmatrix}
  0 & -1 \\
  -(\xi+w_1-u_1) & 0
\end{pmatrix}-
\begin{pmatrix}
  0 & -1 \\
  -(\xi+w_1-u_1) & 0
\end{pmatrix}
\begin{pmatrix}
  V_\xi & U_\xi \\
  W_\xi & -V_\xi
\end{pmatrix}=\\[7pt]
=\begin{pmatrix}
  -U_\xi(\xi+w_1-u_1) & -V_\xi \\
  V_\xi(\xi+w_1-u_1) & -W_\xi
\end{pmatrix}-
\begin{pmatrix}
  -W_\xi & V_\xi \\
  -V_\xi(\xi+w_1-u_1) & -U_\xi(\xi+w_1-u_1)
\end{pmatrix}=\\[7pt]
=\begin{pmatrix}
  W_\xi-U_\xi(\xi+w_1-u_1) & -2V_\xi \\
  2V_\xi(\xi+w_1-u_1) & -W_\xi+U_\xi(\xi+w_1-u_1)
\end{pmatrix}=
\begin{pmatrix}
  \dot{V}_\xi & \dot{U}_\xi \\
  \dot{W}_\xi & -\dot{V}_\xi
\end{pmatrix}=
\dot{L}_\xi.
\end{multline*}
\end{proof}

\section{Integrals}\label{sect4}

\subsection{Integrals of the Neumann--Moser system}\label{sect4.1} \text{}\\
From Theorem \ref{th2}, the Hamiltonian of the Neumann--Moser system \eqref{f-23}--\eqref{f-25} is
\begin{equation*}
H_\xi=H_\xi(U,V,W)=U_\xi W_\xi+V_\xi^2 = \xi^{2n+1} + \sum_{i=1}^{2n+1}h_i \xi^{2n+1-i}.
\end{equation*}
For the sake of completeness, let us check independently that
\begin{lemma}
$H_\xi(U,V,W)$ depends only on parameters $a_1,\,\ldots,\,a_{n+1}$.
\end{lemma}
\begin{proof}
\begin{multline*}
(V_\xi^2+U_\xi W_\xi)' =2V_\xi V_\xi'+U_\xi' W_\xi+W_\xi'U_\xi =\\
=2V_\xi \big(-(\xi+w_1-u_1)U_\xi + W_\xi\big) - 2V_\xi W_\xi + 2(\xi+w_1-u_1)V_\xi U_\xi =0.
\end{multline*}
\end{proof}
Therefore, $h_i$ are constant for $1\leq i\leq 2n+1$ and fixed parameters $a_1,\,\ldots,\,a_{n+1}$.
\begin{theorem}\label{h_theorem}
The Neumann--Moser system \eqref{f-23}--\eqref{f-25} has $2n+1$ integrals
 \begin{align}
    h_1 &= w_1+u_1, \label{h-1} \\[2pt]
    h_2 &=w_1u_1+w_2+u_2, \label{h-2}\\[2pt]
    h_k &=\sum_{i=1}^{k-2}(u_i w_{k-i}+v_i v_{k-1-i})+w_{1}u_{k-1}+u_k+w_k\quad \textrm{if }\; 3\leq k\leq n, \label{h-i-1}\\[2pt]
    h_{n+1}&=\sum_{i=1}^{n-1}(u_i w_{n+1-i}+v_i v_{n-i})+w_{1}u_{n}+w_{n+1},\\[2pt]
    h_k &= \sum_{i=k-1-n}^{n}(u_i w_{k-i}+v_i v_{k-1-i})\quad \textrm{if }\, n+2\leq k\leq 2n+1. \label{h-i-2}
  \end{align}
  \end{theorem}
\begin{proof}
Let us compute
\begin{align*}
V^2_\xi &= \Big(\sum_{i=1}^{n}v_i\xi^{n-i}\Big)^2=
\sum_{l=2}^{2n}\sum_{\substack{i+j=l\\1\leq i,j \leq n}}v_i v_{j}\xi^{2n-l}=
\sum_{k=3}^{2n+1}\sum_{\substack{i+j=k-1\\1\leq i,j \leq n}}v_i v_{j}\xi^{2n+1-k},\\
U_\xi W_\xi &= \Big(\xi^n\, +\, \sum_{i=1}^{n}u_i\xi^{n-i}\Big)\Big(\xi^{n+1} + \sum_{i=1}^{n+1}w_i\xi^{n+1-i}\Big)=\\[2pt]
&=\xi^{2n+1} + \sum_{i=1}^{n}u_i\xi^{2n+1-i}+\sum_{i=1}^{n+1}w_i\xi^{2n+1-i}+\sum_{i=1}^{n}\sum_{j=1}^{n+1}u_i w_j\xi^{2n+1-i-j}=\\
&=\xi^{2n+1} + \sum_{i=1}^{n}(u_i+w_i)\xi^{2n+1-i}+w_{n+1}\xi^n+\sum_{k=2}^{2n+1}\sum_{\substack{i+j=k\\1\leq i \leq n\\ 1\leq j\leq n+1}}u_i w_j\xi^{2n+1-k}.%
\end{align*}
Consider
\begin{align*}
&\sum_{k=3}^{2n+1}\Big(\sum_{\substack{i+j=k\\1\leq i \leq n\\ 1\leq j\leq n+1}}u_i w_j+
\sum_{\substack{i+j=k-1\\1\leq i,j \leq n}}v_i v_{j}\Big)\xi^{2n+1-k}=\\
&=
\sum_{k=3}^{n+1}\Big(\!\!\sum_{\substack{i+j=k\\1\leq i \leq n\\ 1\leq j\leq n+1}}u_i w_j+\!\!
\sum_{\substack{i+j=k-1\\1\leq i,j \leq n}}v_i v_{j}\Big)\xi^{2n+1-k}+
\sum_{k=n+2}^{2n+1}\Big(\!\!\sum_{\substack{i+j=k\\1\leq i \leq n\\ 1\leq j\leq n+1}}u_i w_j+\!\!
\sum_{\substack{i+j=k-1\\1\leq i,j \leq n}}v_i v_{j}\Big)\xi^{2n+1-k}=\\
&=
\sum_{k=3}^{n+1}\Big(u_{k-1}w_{1}+\sum_{i=1}^{k-2}(u_i w_{k-i}+v_i v_{k-1-i})\Big)\xi^{2n+1-k}+
\sum_{k=n+2}^{2n+1}\sum_{i=k-1-n}^{n}(u_i w_{k-i}+v_i v_{k-1-i})\xi^{2n+1-k}.
\end{align*}
Therefore
\begin{align*}
&H_\xi =\xi^{2n+1} + \sum_{i=1}^{n}(u_i+w_i)\xi^{2n+1-i}+w_{n+1}\xi^n+u_1w_1\xi^{2n-1}+\\
&+\sum_{k=3}^{n+1}\Big(u_{k-1}w_{1}+\sum_{i=1}^{k-2}(u_i w_{k-i}+v_i v_{k-1-i})\Big)\xi^{2n+1-k}+
\sum_{k=n+2}^{2n+1}\sum_{i=k-1-n}^{n}(u_i w_{k-i}+v_i v_{k-1-i})\xi^{2n+1-k}=\\
&=\xi^{2n+1} + (u_1+w_1)\xi^{2n}+(u_2+w_2+u_1w_1)\xi^{2n-1}+\\
&+\sum_{k=3}^{n}\big(u_k+w_k+\sum_{i=1}^{k-2}(u_i w_{k-i}+v_i v_{k-1-i})+u_{k-1}w_{1}\big)\xi^{2n+1-k}+\\
&+\Big(\sum_{i=1}^{n-1}(u_i w_{n+1-i}+v_i v_{n-i})+u_{n}w_{1}+w_{n+1}\Big)\xi^{n}+\sum_{k=n+2}^{2n+1}\sum_{i=k-1-n}^{n}(u_i w_{k-i}+v_i v_{k-1-i})\xi^{2n+1-k}.
\end{align*}
\end{proof}

\begin{corollaryt}$\Gamma(t)=h_1-2u_1.$
\end{corollaryt}

\subsection{Integrals of the Neumann system}\label{sect4.2} \text{}\\
We would like to remind the reader of the following classical result of Uhlenbeck \cite{Uhlenbeck-75} and Devaney \cite{Devaney-78}:
\begin{lemma}
The Neumann system has $n+1$ integrals of the form
\[
F_j=x_j^2+\sum_{1\leq k \leq n+1}^{k\neq j}\frac{(x_ky_j-x_jy_k)^2}{a_j-a_k},\quad 1 \leq j\leq n+1.
\]
\end{lemma}
For the sake of convenience and completeness, let us present a proof of this lemma.
\begin{proof}
Let us compute $H_\xi$ in terms of $A,\,X,\,Y$. Since
\[
V_\xi^2=\Big(\sqrt{-1}f_\xi\sum_{j=1}^{n+1}\frac{x_jy_j}{\xi-a_j}\Big)^2=
f_\xi^2\Big(-\sum_{j=1}^{n+1}\frac{x_j^2y_j^2}{(\xi-a_j)^2}
-\sum_{1\leq i< j \leq n+1}\frac{2x_iy_i x_jy_j}{(\xi-a_i)(\xi-a_j)}\Big)
\]
and
\begin{multline*}
U_\xi W_\xi = f_\xi\sum_{j=1}^{n+1}\frac{x_j^2}{\xi-a_j}f_\xi\sum_{j=1}^{n+1}\Big(\frac{y_j^2}{\xi-a_j}+1\Big)=\\
= f_\xi^2\Big(\sum_{1\leq i< j \leq n+1}\frac{x_i^2y_j^2+x_j^2y_i^2}{(\xi-a_i)(\xi-a_j)} + \sum_{j=1}^{n+1}\frac{x_j^2y_j^2}{(\xi-a_j)^2}
+\sum_{j=1}^{n+1}\frac{x_j^2}{\xi-a_j}\Big),
\end{multline*}
one obtains
$$
V_\xi^2+U_\xi W_\xi=f_\xi^2\Big(\sum_{j=1}^{n+1}\frac{x_j^2}{\xi-a_j}+\sum_{1\leq i< j \leq n+1}\frac{(x_iy_j-x_jy_i)^2}{(\xi-a_i)(\xi-a_j)}\Big).
$$
Since
$$
\frac{1}{(\xi-a_i)(\xi-a_j)}=
\frac{1}{a_j-a_i}\Big(\frac{1}{\xi-a_j}-\frac{1}{\xi-a_i}\Big)=
\frac{1}{a_j-a_i}\frac{1}{\xi-a_j}+\frac{1}{a_i-a_j}\frac{1}{\xi-a_i},
$$
one gets
\[
V_\xi^2+U_\xi W_\xi=f_\xi^2 \sum_{j=1}^{n+1}\frac{1}{\xi-a_j}\Big(x_j^2+\sum_{1\leq k  \leq n+1}^{ k\neq j}\frac{(x_ky_j-x_jy_k)^2}{a_j-a_k}\Big)=
f_\xi\sum_{j=1}^{n+1}F_j\prod_{1\leq k\leq n+1}^{k\neq j}(\xi-a_k),
\]
where
\[
F_j=x_j^2+\sum_{1\leq k \leq n+1}^{k\neq j}\frac{(x_ky_j-x_jy_k)^2}{a_j-a_k}.
\]

Denote
\begin{equation}\label{Hhat}
\widehat{H}_\xi=\xi^{n}+\sum_{i=1}^{n}\hat{h}_i\xi^{n-i}=\frac{H_\xi}{f_\xi}=\sum_{j=1}^{n+1}F_j\prod_{1\leq k\leq n+1}^{k\neq j}(\xi-a_k).
\end{equation}
The following two lemmas are obtained by simple direct calculations.
\begin{lemma}
$$\sum_{j=1}^{n+1}F_j=\sum_{j=1}^{n+1}x_j^2=1.$$
\end{lemma}
\begin{lemma}\label{F-h_hat-lemma}
Let $M^{(1,1)}_A$ be the submatrix of $M_A$ (see Equation \eqref{M_A}) formed by deleting the first row and the first column.
Then
$$(1,\;\hat{h}_1,\;\ldots,\;\hat{h}_{n})^T=
M^{(1,1)}_A\cdot (F_1,\;\ldots,\;F_{n+1})^T$$
and
$$(F_1,\;\ldots,\;F_{n+1})^T=
\big(M^{(1,1)}_A\big)^{-1}\cdot (1,\;\hat{h}_1,\;\ldots,\;\hat{h}_{n})^T.$$
\end{lemma}
We know that $\dot{\widehat{H}}_\xi=\dot{H}_\xi/f_\xi=0$, therefore $\dot{\hat{h}}_i=0$ for all $1\leq i\leq n+1$. Therefore, for all $1\leq j\leq n+1$, we have $\dot{F}_j=0$, meaning that $F_j$ is an integral of the Neumann system.
\end{proof}

\subsection{Transformation of integrals}\label{sect4.3} \text{}\\
A natural question regarding the integrals $h_1$, $\ldots$, $h_{2n+1}$ of the Neumann--Moser system \eqref{f-23}--\eqref{f-25} is whether all of them are independent. It turns out that there exists a linear bijection between vectors $H^{\mathrm{I}}=(1,\,h_1,\,\ldots,\,h_{n})$ and $H^{\mathrm{II}}=(h_{n+1},\,\ldots,\,h_{2n+1})$.
\begin{lemma} \label{D12-lem}
There exist invertible $(n+1)\times (n+1)$-matrices $D^{\mathrm{I}}_A,$ $D^{\mathrm{II}}_A$ (see below) such that
\begin{align*}
  H^{\mathrm{I}}&=(1,\;h_1,\;\ldots,\;h_{n})^T=D^{\mathrm{I}}_A\cdot   (1,\;\hat{h}_1,\;\ldots,\;\hat{h}_{n})^T,\\
  H^{\mathrm{II}}&=(h_{n+1},\;h_{n+2},\;\ldots,\;h_{2n+1})^T=D^{\mathrm{II}}_A\cdot   (1,\;\hat{h}_1,\;\ldots,\;\hat{h}_{n})^T.
\end{align*}
\end{lemma}
\begin{proof}
Using the generating polynomial $\widehat{H}_\xi$ defined by Equation \eqref{Hhat}, one can write
$$\xi^{2n+1} + \sum_{i=1}^{2n+1}h_i \xi^{2n+1-i}=H_\xi=
f_\xi\widehat{H}_\xi=
\Big(\xi^{n+1}+\sum_{j=1}^{n+1}e_j\xi^{n+1-j}\Big)\Big(\xi^{n}+\sum_{i=1}^{n}\hat{h}_i\xi^{n-i}\Big).$$
It is easy to see that
\begin{align*}
  (1,\;h_1,\;\ldots,\;h_{n})^T&=D^{\mathrm{I}}_A\cdot   (1,\;\hat{h}_1,\;\ldots,\;\hat{h}_{n})^T,\\
  (h_{n+1},\;h_{n+2},\;\ldots,\;h_{2n+1})^T&=D^{\mathrm{II}}_A\cdot   (1,\;\hat{h}_1,\;\ldots,\;\hat{h}_{n})^T,
\end{align*}
where $D^{\mathrm{I}}_A=D^{\mathrm{I}}(-A)$ and $D^{\mathrm{II}}_A=D^{\mathrm{II}}(-A)$ are $(n+1)\times (n+1)$-matrices
\begin{equation*}D^{\mathrm{I}}_A=\begin{pmatrix}
  1 & 0 & 0 & \ldots & 0 & 0\\
  e_1 & 1 & 0 & \ldots & 0 & 0\\
  e_2 & e_1 & 1 & \ldots & 0 & 0\\
 \vdots & \vdots & \vdots & \ddots & \vdots & \vdots \\
  e_{n-1} & e_{n-2} & e_{n-3} &\ldots & 1 & 0\\
  e_{n} & e_{n-1}  &e_{n-2} & \ldots  & e_1 & 1
  \end{pmatrix}, \quad \quad
  D^{\mathrm{II}}_A=
  \begin{pmatrix}
  e_{n+1} & e_{n} & e_{n-1} &  \ldots & e_2 & e_1 \\
  0& e_{n+1} & e_{n} & \ldots  & e_3 & e_2  \\
  0& 0 & e_{n+1} & \ldots  & e_4 & e_3  \\
  \vdots & \vdots & \vdots & \ddots  & \vdots & \vdots \\
  0 & 0 & 0 & \ldots & e_{n+1} & e_{n}\\
  0 & 0 & 0 & \ldots & 0 & e_{n+1}
  \end{pmatrix}.
\end{equation*}
Here, $e_k=e_k(-A)$,\; $e_k^{(l)}=e_k^{(l)}(-A)$ stand for the elementary symmetric polynomials in variables $-a_1,\,\ldots,\,-a_n$. For the formal definition, please refer to Definition \ref{symmetricdef1} in Appendix A.
\end{proof}
Therefore,
\begin{theorem}Let $h_1,\,\ldots,\,h_{2n+1}$ be the integrals of the Neumann--Moser system \eqref{f-23}--\eqref{f-25} (see Theorem \ref{h_theorem}). Then
\begin{equation*}
  H^{\mathrm{II}}=(h_{n+1},\;h_{n+2},\;\ldots,\;h_{2n+1})^T=D^{\mathrm{II}}_A\cdot\big(D^{\mathrm{I}}_A\big)^{-1} \;\cdot   (1,\;h_1,\;\ldots,\;h_{n})^T=D^{\mathrm{II}}_A\cdot\big(D^{\mathrm{I}}_A\big)^{-1} \cdot H^{\mathrm{I}}.
\end{equation*}
\end{theorem}
\begin{proof}
Both $D^{\mathrm{I}}_A$ and $D^{\mathrm{II}}_A$ are obviously invertible. The inverse of $D^{\mathrm{I}}_A$ is
\begin{equation*}
 \big(D^{\mathrm{I}}_A\big)^{-1}= \begin{pmatrix}
  1 & 0 & \ldots & 0 & 0 & 0\\
  E_1 & 1 & \ldots & 0 & 0 & 0\\
  E_2 & E_1 & \ldots & 0 & 0 & 0\\
 \vdots & \vdots & \ddots & \vdots & \vdots & \vdots \\
 E_{n-1} & E_{n-2} & \ldots & E_1 & 1 & 0\\
 E_{n}   & E_{n-1} & \ldots & E_2 & E_1 & 1
\end{pmatrix},
\end{equation*}
where $E_k=E_k(A)$ for $1\leq k \leq n$ stand for complete homogeneous symmetric polynomials in variables $a_1,\,\ldots,\,a_{n+1}$ (note that the polynomials $e_k$ in $D^{\mathrm{I}}_A$ and $D^{\mathrm{II}}_A$ still stand for $e_k(A)$ in $-a_1,\,\ldots,\,-a_{n+1}$). For the formal definition, please refer to Definition \ref{symmetricdef1-1} in Appendix A. For proof, see Lemma \ref{inverse_D_lemma} in Appendix A.

It is obvious that
\begin{align*}
  (1,\;\hat{h}_1,\;\ldots,\;\hat{h}_{n})^T
  &=\big(D^{\mathrm{I}}_A\big)^{-1} \cdot (1,\;h_1,\;\ldots,\;h_{n})^T=\\
  &=\big(D^{\mathrm{II}}_A\big)^{-1}\cdot (h_{n+1},\;h_{n+2},\;\ldots,\;h_{2n+1})^T.
\end{align*}
The statement of the theorem follows.
\end{proof}

The integrals $h_1,\,\ldots,\,h_{2n+1}$ can be rewritten as functions in $A,\,X,\,Y$, giving explicit integrals of the Neumann system. For example, it is easy to see that
$$h_1=\sum_{k=1}^{n+1}a_k(x_k^2-1)+\sum_{k=1}^{n+1}(y_k^2-a_k)=\sum_{k=1}^{n+1}\big(y^2_k+a_kx_k^2-2a_k\big).$$
Indeed, one can check independently that
$$\dot{h}_1=\sum_{k=1}^{n+1}\big(-2y_kx_k(\Gamma +a_k)+2a_kx_ky_k\big)=-2g\sum_{k=1}^{n+1}y_kx_k=0.$$
The case of $h_i$ for bigger $i$ is considerably more difficult. To tackle it, we want to establish explicitly the relationship between the integrals $F_1,\;\ldots,\;F_{n+1}$ of the Neumann system and the integrals $h_1,\;\ldots,\;h_{2n+1}$ of the Neumann--Moser system.

\begin{theorem}\label{F-h_th}The integrals $F_1,\,\ldots,\,F_{n+1}$ of the Neumann system and the integrals $h_1,\,\ldots,\,h_{2n+1}$ of the Neumann--Moser system are related by
\begin{align*}
  (F_1,\;\ldots,\;F_{n+1})^T
  &=\big(D^{\mathrm{I}}_A\cdot M^{(1,1)}_A\big)^{-1}\cdot (1,\;h_1,\;\ldots,\;h_{n})^T=\\
  &=\big(D^{\mathrm{II}}_A \cdot M^{(1,1)}_A\big)^{-1}\cdot (h_{n+1},\;h_{n+2},\;\ldots,\;h_{2n+1})^T,
\end{align*}
where $M^{(1,1)}_A$ is the submatrix of $M_A$ (see Equation \eqref{M_A}) formed by deleting the first row and the first column
and $D^{\mathrm{I}}_A,$ $D^{\mathrm{II}}_A$ are given by Lemma \ref{D12-lem}.
\end{theorem}
\begin{proof}
From Lemma \ref{D12-lem} and Lemma \ref{F-h_hat-lemma},
 \begin{align*}
  (1,\;h_1,\;\ldots,\;h_{n})^T&=D^{\mathrm{I}}_A\cdot M^{(1,1)}_A \;\cdot   (F_1,\;\ldots,\;F_{n+1})^T,\\
  (h_{n+1},\;h_{n+2},\;\ldots,\;h_{2n+1})^T&=D^{\mathrm{II}}_A\cdot M^{(1,1)}_A \;\cdot   (F_1,\;\ldots,\;F_{n+1})^T.
\end{align*}
Thus,
\begin{align*}
  (F_1,\;\ldots,\;F_{n+1})^T
  &=\big(D^{\mathrm{I}}_A\cdot M^{(1,1)}_A\big)^{-1}\cdot (1,\;h_1,\;\ldots,\;h_{n})^T=\\
  &=\big(D^{\mathrm{II}}_A\cdot M^{(1,1)}_A\big)^{-1}\cdot (h_{n+1},\;h_{n+2},\;\ldots,\;h_{2n+1})^T.
\end{align*}
\end{proof}

\section{Recurrent Solutions of the Neumann--Moser system}\label{sect5}
Using Theorem \ref{h_theorem} and Corollary \ref{vw-u_cor}, one can obtain recurrent relations for $u_1,\,\ldots,\,u_n$ that do not contain $v_1,\,\ldots,\,v_n$ and $w_1,\,\ldots,\,w_{n+1}$.
\begin{corollaryt}\label{u-rec-cor-1}\text{}\\
From Equations \eqref{g-formula} and \eqref{h-1},\quad $u_1=\frac{1}{2}(h_1-\Gamma)$ for all $n\geq 1$.\\[4pt]
From Corollary \ref{vw-u_cor} and Equation \eqref{h-2},
\[
u_2=\frac{1}{4}\ddot{u}_1+\frac{3}{2}u_1^2-h_1u_1+\frac{1}{2}h_2=
-\frac{1}{8}(\ddot{\Gamma}-3\Gamma^2+10h_1\Gamma-7h_1^2-4h_2).
\]
\end{corollaryt}

\begin{theorem}\label{u-rec-th}
For $n\geq 3$ and for all $3\leq k\leq n$, one has
\begin{multline}\label{u-rec}
  u_k\,=\,\frac{1}{4}\ddot{u}_{k-1}-(h_1-\frac{3}{2}u_1)u_{k-1}+\frac{1}{2}h_k-\\
  -\frac{1}{8}\sum_{i=1}^{k-2}\Big(4u_iu_{k-i}+\dot{u}_{i}\dot{u}_{k-i-1}-2u_i\ddot{u}_{k-i-1}+4(h_1-2u_1)u_iu_{k-1-i}\Big).
\end{multline}
\end{theorem}
\begin{proof}
From Corollary \ref{vw-u_cor} and Equation \eqref{h-i-1}, for all $3 \leq k\leq n$,
\begin{equation*}
w_k=
  \begin{cases}
      u_k-(u_1-w_1)u_{k-1}+\dot{v}_{k-1}=u_k+(h_1-2u_1)u_{k-1}+\frac{1}{2}\ddot{u}_{k-1},\\
h_k-u_k-\sum_{i=1}^{k-1}u_iw_{k-i}-\sum_{i=1}^{k-2}v_{i}v_{k-1-i}=
h_k-u_k-\sum_{i=1}^{k-1}u_iw_{k-i}-\frac{1}{4}\sum_{i=1}^{k-2}\dot{u}_{i}\dot{u}_{k-1-i}.
  \end{cases}
\end{equation*}
Therefore,
\begin{align}
  u_k&\,=\,\;\,\;\frac{1}{4}\ddot{u}_{k-1}-\frac{1}{2}(h_1-2u_1)u_{k-1}-\frac{1}{2}h_k
  -\frac{1}{8}\sum_{i=1}^{k-2}\dot{u}_{i}\dot{u}_{k-1-i}-\frac{1}{2}\sum_{i=1}^{k-1}u_iw_{k-i}, \label{PQ1}\\
  w_k&\,=\,-\frac{1}{4}\ddot{u}_{k-1}+\frac{1}{2}(h_1-2u_1)u_{k-1}-\frac{1}{2}h_k
  -\frac{1}{8}\sum_{i=1}^{k-2}\dot{u}_{i}\dot{u}_{k-1-i}-\frac{1}{2}\sum_{i=1}^{k-1}u_iw_{k-i}. \label{PQ2}
\end{align}
Note that
\begin{multline*}
-\frac{1}{8}\sum_{i=1}^{k-2}\dot{u}_{i}\dot{u}_{k-1-i}-\frac{1}{2}\sum_{i=1}^{k-1}u_iw_{k-i}=\\
=-\frac{1}{8}\sum_{i=1}^{k-2}\Big(4u_iu_{k-i}+\dot{u}_{i}\dot{u}_{k-i-1}-2u_i\ddot{u}_{k-i-1}+4(h_1-2u_1)u_iu_{k-1-i}\Big)-
\frac{1}{2}u_{k-1}(h_1-u_1).
\end{multline*}
The Equation \eqref{u-rec} follows.\end{proof}

\begin{corollaryt}\label{u-rec-cor-2}
Let $(U,\,V,\,W)$ be a solution of the Neumann--Moser dynamical system. Then the functions $u_1,\,\ldots,\,u_n$ form the recursion given by Corollary \ref{u-rec-cor-1} and Theorem \ref{u-rec-th}, and
\begin{equation}\label{f-23-1}
\ddot{u}_{k-1}\,=\,4(h_1-\frac{3}{2}u_1)u_{k-1}-2h_k
  +\frac{1}{2}\sum_{i=1}^{k-2}\Big(4u_iu_{k-i}+\dot{u}_{i}\dot{u}_{k-i-1}-2u_i\ddot{u}_{k-i-1}+4(h_1-2u_1)u_iu_{k-1-i}\Big).
\end{equation}
Moreover, Equation \eqref{f-23-1} is an ordinary differential equation in the function $u_1(t)$ of order $2n$ with $n+1$ free parameters $h_{1},\,\ldots,\,h_{n+1}$.
\end{corollaryt}

\section{From the Neumann--Moser system to the Mumford system and back}\label{sect6}
In 1984, Mumford \cite[\S 3]{Mumford-84} introduced a certain dynamical system that was later named in his honor by Vanhaecke, who developed and generalized its theory (see \cite[Chapter VI]{Vanhaecke-96}). This system has been used extensively in connection with several topics in the realm of integrable systems, including KP vector fields, Sato Grassmanians, the Krichiver map, and the Korteweg--De Vries hierarchy.

We will follow Buchstaber \cite[Section 1]{Buch23} in our notations and definitions.
\begin{definition}
Consider the space $\mathbb{C}^{n}$, where $n\geqslant 1$, with the vector of coordinates $\mathbf{t}=(t_1,\,\ldots,\,t_n)$. Define the differentiation operators $\mathcal{D}_\eta = \sum_{i=1}^n \eta^{n-i}\partial_i$, where $\eta$ is a formal (``generating'') variable and $\partial_i = \partial/\partial t_i$. Note that, for independent parameters $\xi$ and $\eta$, one has $\mathcal{D}_\eta\xi = \mathcal{D}_\xi\eta \equiv 0$.

Consider mappings
\[
L_\xi \colon \mathbb{C}^{n}\to \mathcal{S}L(2,\mathbb{C}[\xi])\quad \text{and}\quad A_\eta \colon \mathbb{C}^{n}\to \mathcal{S}L(2,\mathbb{C}[\eta])
\]
such that
$$L_\xi(\mathbf{t}) =
\begin{pmatrix}
  V_\xi(\mathbf{t}) & U_\xi(\mathbf{t}) \\
  W_\xi(\mathbf{t}) & -V_\xi(\mathbf{t})
\end{pmatrix},
\quad\quad
A_\eta(\mathbf{t}) = \begin{pmatrix}
  0 & 0 \\
  U_\eta(\mathbf{t}) & 0
\end{pmatrix},$$
where
\begin{equation}\label{f-1-1}
U_\xi(\mathbf{t}) = \xi^n+\sum_{i=1}^n u_i(\mathbf{t})\xi^{n-i},\quad
V_\xi(\mathbf{t}) = \sum_{i=1}^n v_i(\mathbf{t})\xi^{n-i},\quad
W_\xi(\mathbf{t}) = \xi^{n+1}+\sum_{i=1}^{n+1} w_i(\mathbf{t})\xi^{n+1-i}.
\end{equation}
Here, $(U;\;V;\;W)=(u_1,\,\ldots,\,u_n;\;v_1,\,\ldots,\,v_n;\;w_1,\,\ldots,\,w_{n+1})$ belongs to $\mathbb{C}^{3n+1}$.

Then
\begin{equation}\label{f-1}
\mathcal{D}_\eta L_\xi(\mathbf{t}) = \frac{1}{\xi-\eta}[L_\xi(\mathbf{t}),L_\eta(\mathbf{t})] + [L_\xi(\mathbf{t}),A_\eta(\mathbf{t})].
\end{equation}
We have obtained system \emph{the Mumford system} in the Lax form. Note that it is quadratic in each of the variables on $\mathbb{C}^{3g+1}$.
\end{definition}
Writing the Mumford system in its expanded form, one obtains the system
\begin{align}\label{f-2}
\mathcal{D}_\eta u_\xi&=\frac{2}{\xi-\eta}(v_\xi u_\eta-u_\xi v_\eta),\\
\label{f-3} \mathcal{D}_\eta v_\xi&=\frac{1}{\xi-\eta}(u_\xi w_\eta - w_\xi u_\eta) + u_\xi u_\eta,\\
\label{f-4} \mathcal{D}_\eta w_\xi&=\frac{2}{\xi-\eta}(w_\xi v_\eta - v_\xi w_\eta) - 2v_\xi u_\eta.
\end{align}
Directly from equations \eqref{f-2}--\eqref{f-4}, one has
\begin{corollaryt}
\label{C-1}
\begin{align}
\mathcal{D}_\xi v_\eta\, &= \, \mathcal{D}_\eta v_\xi,\\
\mathcal{D}_\xi u_\eta\, &= \, \mathcal{D}_\eta u_\xi\, = \,\frac{\mathcal{D}_\xi w_\eta - \mathcal{D}_\eta w_\xi}{\xi-\eta}.
\end{align}
\end{corollaryt}

\begin{theorem}\label{M-NS-th}
Let
$
\big(U;\;V;\;W\big)\,=\,
\big(u_1,\,\ldots,\,u_n;\;
 v_1,\,\ldots,\,v_n;\;
 w_1,\,\ldots,\,w_{n+1}\big)
$
be $3g+1$ functions in $\mathbf{t}=(t_1,\,\ldots,\,t_n)$. These functions solve the Mumford dynamical system if and only if they solve the Neumann--Moser dynamical system with respect to the differentiation operator $\partial=\partial_1$ in $t=t_1$.
\end{theorem}

\begin{proof}
The recurrence relations on $u_k$ and $w_k$ given by Equations \eqref{PQ1}--\eqref{PQ2} coincide with the recurrent solutions of the Mumford system given by \cite[Corollary 4.4, also Equations (7), (8)]{Buch23}. Therefore, if $(U;\;V;\;W)$ solve the Neumann--Moser dynamical system with respect to $\partial=\partial_1$, then they solve the Mumford dynamical system.

Conversely, suppose that $(U;\;V;\;W)$ solve the Mumford system. Note that $\partial_i$ is the coefficient of $\eta^{n-1}$ in $\mathcal{D}_\eta = \sum_{i=1}^n \eta^{n-i}\partial_i$. Therefore, one can isolate the dependence of the Mumford system on the time variable $t=t_1$ by computing the coefficients of $\eta^{n-1}$ in $\mathcal{D}_\eta u_\xi$, $\mathcal{D}_\eta v_\xi$, and $\mathcal{D}_\eta w_\xi$ (treated as polynomials in $\eta$) respectively. It remains to prove Lemma \ref{M-NS-lem} below in order to obtain the statement of the theorem.\end{proof}

\begin{lemma}\label{M-NS-lem}
Consider $\mathcal{D}_\eta u_\xi$, $\mathcal{D}_\eta v_\xi$, and $\mathcal{D}_\eta w_\xi$ as polynomials in $\eta$. Then
\begin{enumerate}
  \item The coefficient of $\eta^{n-1}$ in $\mathcal{D}_\eta u_\xi$ is $\;-2V_\xi$.
  \item The coefficient of $\eta^{n-1}$ in $\mathcal{D}_\eta v_\xi$ is $\;-(\xi+w_1-u_1)U_\xi + W_\xi$.
  \item The coefficient of $\eta^{n-1}$ in $\mathcal{D}_\eta w_\xi$ is $\;2(\xi+w_1-u_1)V_\xi$.
\end{enumerate}
\end{lemma}
\begin{proof}
For the sake of convenience, set introduce $\zeta^{k,l} = \frac{\xi^k\eta^l-\eta^k\xi^l}{\xi-\eta}$. Note that $\zeta^{k,l} = -\zeta^{l,k}$,\, $\zeta^{k,k} = 0$,\, and $\zeta^{k+1,k} = \xi^k\eta^k$. Moreover, for $k>l$, one has
\begin{equation}\label{ff-1}
\zeta^{k,l} = \sum_{s=0}^{k-l-1}\xi^{k-1-s}\eta^{l+s}=-\zeta^{l,k}.
\end{equation}

\textbf{(1)} First, consider $\mathcal{D}_\eta u_\xi$:
  \begin{multline*}
 \mathcal{D}_\eta u_\xi=\frac{2}{\xi-\eta}(v_\xi u_\eta-u_\xi v_\eta)=\\
 =\frac{2}{\xi-\eta}\Big(
 \sum_{j=1}^n v_j\xi^{n-j}\big(\eta^n+\sum_{i=1}^n u_i\eta^{n-i}\big)-
\big(\xi^n+\sum_{i=1}^n u_i\xi^{n-i}\big) \sum_{j=1}^n v_j\eta^{n-j}
 \Big) =\\
 =\frac{2}{\xi-\eta}\Big(\sum_{j=1}^n v_j(\xi^{n-j}\eta^n-\xi^n\eta^{n-j})+
 \sum_{i=1}^n\sum_{j=1}^n u_iv_j (\xi^{n-j}\eta^{n-i}-\xi^{n-i}\eta^{n-j})
 \Big) =  \\
=2\!\sum_{j=1}^n v_j\zeta^{n-j,n}\!+2\!\sum_{i=1}^n\!\sum_{j=1}^n u_iv_j\zeta^{n-j,n-i}\!
=-2\!\sum_{j=1}^n v_j\zeta^{n,n-j}\!+2\!\sum_{i=1}^n\!\sum_{j=1}^{i-1} (u_iv_j-u_jv_i)\zeta^{n-j,n-i}\!=\\
=-2\sum_{j=1}^n v_j\sum_{s=0}^{j-1}\xi^{n-1-s}\eta^{n-j+s}+2\sum_{i=1}^n\sum_{j=1}^{i-1} (u_iv_j-u_jv_i)\sum_{s=0}^{i-j-1}\xi^{n-1-j-s}\eta^{n-i+s}.
\end{multline*}
Then the coefficient of $\eta^{n-1}$ in $\mathcal{D}_\eta u_\xi$ is
$$
-2\sum_{j=1}^n v_j\xi^{n-j}=-2V_\xi.
$$

\textbf{(2)} Second, consider $\mathcal{D}_\eta v_\xi$.  Likewise,
\begin{multline*}
\mathcal{D}_\eta v_\xi=\frac{1}{\xi-\eta}(u_\xi w_\eta - w_\xi u_\eta) + u_\xi u_\eta
=\big( \xi^n+\sum_{i=1}^n u_i \xi^{n-i}\big) \big(\eta^n+\sum_{j=1}^n u_j\eta^{n-j}\big)+\\
+\frac{1}{\xi-\eta}\Big(\!
\big( \xi^n+\sum_{i=1}^n u_i \xi^{n-i}\big)  \big(\eta^{n+1}+\sum_{j=1}^{n+1} w_j\eta^{n+1-j}\big)-
\big(\eta^n+\sum_{i=1}^n u_i\eta^{n-i}\big)  \big( \xi^{n+1}+\sum_{j=1}^{n+1} w_j \xi^{n+1-j}\big)\!\Big)=\\
=\xi^n\eta^n + \sum_{i=1}^n u_i(\xi^n\eta^{n-i}+\xi^{n-i}\eta^{n})+ \sum_{i=1}^n \sum_{j=1}^n u_iu_j \xi^{n-i}\eta^{n-j}+\\
+\zeta^{n,n+1}+\sum_{i=1}^n u_i \zeta^{n-i,n+1}+\sum_{j=1}^{n+1} w_j \zeta^{n,n+1-j}+\sum_{i=1}^n \sum_{j=1}^{n+1}u_iw_j \zeta^{n-i,n+1-j},
\end{multline*}
where
\begin{equation*}
\zeta^{n,n+1}=-\zeta^{n+1,n}=\sum_{s=0}^{2n}\xi^{n-s}\eta^{n+s};\\
\end{equation*}
\begin{align*}
\sum_{i=1}^n u_i \zeta^{n-i,n+1}&=-\sum_{i=1}^n u_i \zeta^{n+1,n-i}
=-\sum_{i=1}^n u_i\sum_{s=0}^{i}\xi^{n-s}\eta^{n-i+s};\\
\sum_{j=1}^{n+1} w_j \zeta^{n,n+1-j}&=\sum_{j=2}^{n+1} w_j \zeta^{n,n+1-j}=\sum_{l=1}^{n} w_{l+1} \zeta^{n,n-l}=
\sum_{l=1}^{n} w_{l+1} \sum_{s=0}^{l-1}\xi^{n-1-s}\eta^{n-l+s};\\
\sum_{i=1}^n \sum_{j=1}^{n+1}u_iw_j \zeta^{n-i,n+1-j}&=-\sum_{j=1}^{n+1}\sum_{i=1}^n u_iw_j \zeta^{n+1-j,n-i}=\\
&=
-\sum_{i=1}^n u_iw_1 \zeta^{n,n-i}-\sum_{l=1}^{n}\sum_{i=1}^n u_iw_{l+1} \zeta^{n-l,n-i}=\\
&=-w_1\sum_{i=1}^n u_i \sum_{s=0}^{i-1}\xi^{n-s-1}\eta^{n-i+s}-\sum_{i=1}^n\sum_{l=1}^{i-1} (u_iw_{l+1}-u_lw_{i+1})\zeta^{n-l,n-i},
\end{align*}
where
\begin{equation*} -\sum_{i=1}^n\sum_{l=1}^{i-1} (u_iw_{l+1}-u_lw_{i+1})\zeta^{n-l,n-i}=
-\sum_{i=1}^n\sum_{l=1}^{i-1} (u_iw_{l+1}-u_lw_{i+1})\sum_{s=0}^{i-l-1}\xi^{n-1-l-s}\eta^{n-i+s}.
\end{equation*}
Then the coefficient of $\eta^{n-1}$ in $\mathcal{D}_\eta v_\xi$ is
\begin{multline*}
u_1\xi^n
+u_1\sum_{i=1}^n u_i \xi^{n-i}
-\sum_{i=1}^n u_i\xi^{n+1-i}
+\sum_{l=1}^{n} w_{l+1} \xi^{n-l}
-w_1\sum_{i=1}^n u_i \xi^{n-i}=\\
=
u_1\big(\xi^n
+\sum_{i=1}^n u_i \xi^{n-i}\big)
-\xi\big(\xi^n
+\sum_{i=1}^n u_i \xi^{n-i}\big)+\xi^{n+1}
+\sum_{j=2}^{n+1} w_{j} \xi^{n+1-j}
-w_1\big(\xi^n
+\sum_{i=1}^n u_i \xi^{n-i}\big)+w_1\xi^n=\\
=
(u_1-w_1-\xi)\big(\xi^n
+\sum_{i=1}^n u_i \xi^{n-i}\big)
+\xi^{n+1}
+\sum_{j=1}^{n+1} w_{j} \xi^{n+1-j}=
-(\xi+w_1-u_1)U_\xi + W_\xi.
\end{multline*}

\textbf{(3)} Third, consider $\mathcal{D}_\eta w_\xi$.  Likewise,
\begin{align*}
\mathcal{D}_\eta w_\xi &=
\frac{2}{\xi-\eta}(w_\xi v_\eta - v_\xi w_\eta) - 2v_\xi u_\eta =\\
&= 2\sum_{j=1}^n v_j\zeta^{n+1,n-j}+2\sum_{i=1}^{n+1}\sum_{j=1}^n w_iv_j\zeta^{n+1-i,n-j}-
2\sum_{j=1}^n v_j\xi^{n-j}\eta^n-2\sum_{i=1}^n\sum_{j=1}^n u_iv_j \xi^{n-j}\eta^{n-i},
\end{align*}
where
\begin{equation*}
2\sum_{j=1}^n v_j\zeta^{n+1,n-j}= 2\sum_{j=1}^n v_j\sum_{s=0}^{j}\xi^{n-s}\eta^{n-j+s}
\end{equation*}
and
\begin{align*}
2\sum_{i=1}^{n+1}\sum_{j=1}^n w_iv_j\zeta^{n+1-i,n-j}&=
2\sum_{j=1}^n w_1v_j\zeta^{n,n-j}+2\sum_{l=1}^{n}\sum_{j=1}^n w_{l+1}v_j\zeta^{n-l,n-j}=\\
&=2w_1\sum_{j=1}^n v_j\sum_{s=0}^{j-1}\xi^{n-1-s}\eta^{n-j+s}+
2\sum_{j=1}^{n}\sum_{l=1}^{j-1} (w_{l+1}v_j-w_{j+1}v_l)\zeta^{n-l,n-j},
\end{align*}
where
\begin{equation*}
2\sum_{j=1}^{n}\sum_{l=1}^{j-1} (w_{l+1}v_j-w_{j+1}v_l)\zeta^{n-l,n-j}
=2\sum_{j=1}^{n}\sum_{l=1}^{j-1} (w_{l+1}v_j-w_{j+1}v_l)\sum_{s=0}^{j-l-1}\xi^{n-1-l-s}\eta^{n-j+s}.
\end{equation*}
Then the coefficient of $\eta^{n-1}$ in $\mathcal{D}_\eta w_\xi$ is
$$2\sum_{j=1}^n v_j\xi^{n+1-j}+2w_1\sum_{j=1}^n v_j-2u_1\sum_{j=1}^n v_j \xi^{n-j}=2(\xi+w_1-u_1)V_\xi.$$
\end{proof}

\section{Connections to the Korteweg--de Vries hierarchy}\label{sect7}

\subsection{The Korteweg--de Vries hierarchy}\label{sect7.1}
\begin{definition}
 The \emph{Korteweg–de Vries hierarchy (KdV hierarchy)} is an infinite sequence of compatible partial differential equations for the function $G =G(t_{0}=t,t_{1};\,t_{2},t_{3},\ldots)$ that has the form
\begin{align}
\label{KdVhierarchy1}
\partial_{k}G_1\,&=\,\partial G_{k+1},\\
\label{KdVhierarchy2}
\partial G_{k+1}\,&=\,\Lambda[G]\,\partial G_{k}
\end{align}
for all $k\in\mathbb{N}$, where $\partial=\frac{\partial}{\partial t}=\frac{\partial}{\partial t_0}$,\, $\partial_{i}=\frac{\partial}{\partial t_i}$, $i\geq 1$, the operator $\Lambda[G]=\frac{1}{4}\partial^{2}-G -\frac{1}{2}\dot{G }\partial^{-1}$ is the pseudo-differential Lenard operator, $G_0$ is a constant, $G_1$ is a function of $G $, $\{G_i\}_{i\geq 2}$ are differential polynomials of $G_1$, and equations \eqref{KdVhierarchy1}--\eqref{KdVhierarchy2} for $k=1$ yield the \emph{Korteweg--de Vries equation (KdV equation)}
\begin{equation*}
4\frac{\partial G }{\partial t_1}=6G \dot{G }-G^{(3)}.
\end{equation*}
We will say that the KdV hierarchy is \emph{$(l+1)$-stationary} if $G_i\equiv 0$ for all $i>l$, but $G_l\neq 0$.
\end{definition}

See, for example, \cite[\S 3.4]{Dunajski-10} and \cite{Newell-87} for more information on the KdV equation and the KdV hierarchy.

\begin{theorem}\label{Lambda-NS-th}
Let
$U(t)=(u_1,\,\ldots,\,u_{n})$, $V(t)=(v_1,\,\ldots,\,v_{n})$,
$W(t)=(w_1,\,\ldots,\,w_{n+1})$
form the Neumann--Moser system \eqref{f-23}--\eqref{f-25}.
Then $u_1,\,\ldots,\,u_n$ are differential polynomials in $\Gamma=w_1-u_1$ such that
\begin{align}
  \dot{u}_1\,&=\,-\frac{1}{2}\dot{\Gamma},\label{dot-u_1}\\
  \dot{u}_{i+1}\,&=\,\frac{1}{4} u_i^{(3)}- \Gamma \dot{u}_i-\frac{1}{2}\dot{\Gamma } u_i\quad \text{ for all}\quad 1\leq i\leq n-1,\label{dot-u_i+1}\\
  0\,&=\,u_n^{(3)} - 4\Gamma \dot{u}_n - 2\dot{\Gamma } u_n. \label{dot-u_n}
\end{align}
\end{theorem}

\begin{proof}
From Equations \eqref{f-23}--\eqref{f-25},
$$
U^{(3)}_\xi - 4(\xi+w_1-u_1)\dot{U}_\xi - 2(\dot{w}_1-\dot{u}_1)U_\xi=0.
$$
Rewriting this in coordinate form, we get
$$
\sum_{i=1}^{n}u^{(3)}_i\xi^{n-i}-4(\xi+w_1-u_1)\sum_{i=1}^{n}\dot{u}_i\xi^{n-i}-2(\dot{w}_1-\dot{u}_1)\Big(\xi^n + \sum_{i=1}^{n}u_i\xi^{n-i}\Big)=0,
$$
\[
\sum_{i=1}^{n}u_i^{(3)}\xi^{n-i}-4\sum_{i=1}^{n}\dot{u}_i\xi^{n+1-i}-4(w_1-u_1)\sum_{i=1}^{n}\dot{u}_i\xi^{n-i}
-2(\dot{w}_1-\dot{u}_1)\xi^n-2(\dot{w}_1-\dot{u}_1)\sum_{i=1}^{n}u_i\xi^{n-i}=0,
\]
\begin{multline*}
(-2\dot{w}_1-2\dot{u}_1)\xi^n+\sum_{i=1}^{n-1}\big(u_i^{(3)}-4\dot{u}_{i+1}-4(w_1-u_1)\dot{u}_i-2(\dot{w}_1-\dot{u}_1)u_i\big)\xi_{n-k}+\\
+(u_n^{(3)}-4(w_1-u_1)\dot{u}_n-2(\dot{w}_1-\dot{u}_1)u_n)=0.
\end{multline*}
Equivalently,
\begin{align*}
  -2(\dot{w}_1+\dot{u}_1) &=0,\\
  u_i^{(3)}-4\dot{u}_{i+1}-4(w_1-u_1)\dot{u}_i-2(\dot{w}_1-\dot{u}_1)u_i &=0\quad \text{for all}\,\,1\leq i\leq n-1,\\
  u_n^{(3)} - 4(w_1-u_1)\dot{u}_n - 2(\dot{w}_1-\dot{u}_1)u_n &=0.
\end{align*}

Rewriting this in $\Gamma=w_1-u_1$, we arrive at Equations \eqref{dot-u_1}--\eqref{dot-u_n}.
\end{proof}

\begin{corollaryt}
  For all $1\leq k\leq n-1$, the functions $\Gamma=h_1-2u_1,\,u_1,\,\ldots,\,u_{n}$ given by the Neumann--Moser system of dimension $n$ satisfy the differential equation
 $$
 \frac{\partial}{\partial t}u_{k+1}\,=\,\Lambda[\Gamma]\,\frac{\partial}{\partial t}u_k,
 \quad \text{where}\quad \Lambda[\Gamma]=\frac{1}{4}\partial^{2}-\Gamma -\frac{1}{2}\dot{\Gamma }\partial^{-1}.
 $$
 Moreover,
 $$\frac{\partial}{\partial t}u_{k+1}\,=\,-\frac{1}{2}\Gamma\quad \text{and}\quad \Lambda[\Gamma]\,\frac{\partial}{\partial t}u_n\,\equiv\,0.$$

\end{corollaryt}

Informally, Theorem \ref{Lambda-NS-th} claims that the functions $u_1,\,\ldots,\,u_n$ from the Neumann--Moser system solve the finite KdV hierarchy of length $n$ for the function $\Gamma$. Below, we give the formal statement of this result.
\begin{theorem}\label{KdV-NS-th}
Let the functions $G=G(t,\,t_{1},\,t_{2},\ldots)$,\, $G_0(G),\,G_1(G),\,\ldots,\,G_{n}(G)$ be such that
\begin{align}
G(t,\,0,\,0,\,0,\,\ldots)&=\Gamma(t),\\
G_0\big(G(t,\,t_{1}\,t_{2},\ldots)\big)&\equiv 1,\\
G_k\big(G(t,\,0,\,0,\,\ldots)\big)&=u_i(t)\quad\quad \;\;\, \text{for all}\quad 1\leq k \leq n,\\
\partial_{k}G_1&=\partial G_{k+1} \quad\quad \text{for all}\quad 1\leq k \leq n-1,\\
\partial_{i}G_1&\equiv 0 \quad\quad\quad\quad\, \text{for all}\quad i \geq n.
\end{align}
Then the functions $G_0(G),\,G_1(G),\,\ldots,\,G_{n}(G)$ form an $(n+1)$-stationary KdV hierarchy for the function $G$ such that $G=h_1-2G_1$ and $$\Lambda[G]\,\partial G_n\,=\,\frac{1}{4}\partial^{(3)}G_n-G\,\partial G_n-\frac{1}{2}G_n\,\partial G \equiv 0.$$
Note that this can be rewritten as a differential equation on $G$ via repeated application of Equation \eqref{KdVhierarchy2}.\\
This hierarchy is dependent on $n$ parameters $h_1,\,\ldots,\,h_{n+1}$ given by Equations \eqref{h-1}-\eqref{h-i-2}.
\end{theorem}

\subsection{The hyperelliptic Kleinian functions}\label{sect7.2} \text{}\\
For the sake of convenience, we quote \cite[Section 8]{Buch23} below. The permission of the author has been obtained.

Consider a $2g$-dimensional family of curves
\[
 V_{\lambda} = \{ (X,Y) \in \mathbb{C}^2\colon Y^2 = F(X)\},
\]
where $F(X) = 4X^{2g+1} + \lambda_4 X^{2g-1} + \cdots + \lambda_{4g-2}$.

Set $\mathcal{D} = \{ \lambda\in\mathbb{C}^{2g}:F(X)\,\text{ has multiple roots}\}$ and $\mathcal{B} = \mathbb{C}^{2g}\setminus \mathcal{D}$.
For each $\lambda\in\mathcal{B}$, we obtain a smooth hyperelliptic curve $\bar{V}_{\lambda}$ of genus $g$
with the Jacobian $\rm{Jac}(\bar{V}_{\lambda}) = \mathbb{C}^g\!/\Gamma_g$.
Here, $\Gamma_g\subset\mathbb{C}^g$ is a lattice of rank $2g$ generated by periods of the integrals of $g$ holomorphic differentials
over $2g$ cycles, which form a basis for the one-dimensional homologies of the curve $\bar{V}_{\lambda}$.

In 1886, Klein posed the problem of constructing hyperelliptic functions of genus $g>1$ that possess properties similar to those of Weierstrass elliptic functions.
In 1898, H.F. Baker addressed Klein's problem for $g=2$, see \cite{Baker-1898}, see also \cite{BEL-19}.
The case of $g>2$ remained an open problem for a long time. It gained significantly more attention with the development of algebraic-geometric methods in soliton theory.

In the paper \cite{BEL-97-2}, it was shown that there exists a unique entire function $\sigma(\mathbf{z};\lambda)$
in $\mathbf{z} = (z_1,\ldots,z_{2g-1})\in \mathbb{C}^g$,
where $\lambda=(\lambda_4,\ldots,\lambda_{4g+2})\in \mathbb{C}^{2g}$,
which is called the \emph{hyperelliptic sigma function}.
In the neighborhood of the point $\mathbf{z}=0$, the coefficients of the series expansion of the function $\sigma(\mathbf{z};\lambda)$ with respect to $\mathbf{z}$ are polynomials in $\lambda$.
The logarithmic derivatives of this function of order $2$ and higher generate the field of meromorphic functions on the Jacobian $\rm{Jac}(\bar{V}_{\lambda})$, called \emph{Abelian functions}. As such, the function $\sigma(\mathbf{z};\lambda)$ is a solution to Klein's problem (see \cite{BEL-19}).

Set $f'(\mathbf{z}) = \frac{\partial}{\partial z_1}f(\mathbf{z})$ and
\[
\wp_{2k} = -\frac{\partial^2}{\partial z_1\partial z_{2k-1}}\ln\sigma(\mathbf{z}), k=1,\ldots,g;\quad
\wp_{2i-1,2k-1} = -\frac{\partial^2}{\partial z_{2i-1}\partial z_{2k-1}}\ln\sigma(\mathbf{z}), i\neq 1,\, k\neq 1.
\]
Functions $\wp_{2k}$, $\wp_{2i-1,2k-1}$, and their derivatives are called the \emph{hyperelliptic Kleininan functions}.
\begin{theorem}[see \cite{BEL-19}]
All polynomial relations in the polynomial ring of the hyperelliptic Kleininan functions are generated by the relations
\begin{equation} \label{f-55}
\wp''_{2i} = 6(\wp_{2i+2} +\wp_{2}\wp_{2i}) - 2(\wp_{3,2i-1} - \lambda_{2i+2}\, \delta_{i,1})
\end{equation}
and
\begin{multline}\label{f-56}
\wp'_{2i}\wp'_{2k} = 4(\wp_{2i}\wp_{2k+2} + \wp_{2i+2}\wp_{2k} + \wp_{2}\wp_{2i}\wp_{2k} + \wp_{2i+1,2k+1}) - \\[3pt]
- 2(\wp_{2i}\wp_{3,2k-1} + \wp_{2k}\wp_{3,2i-1} + \wp_{2i-1,2k+3} +\wp_{2i+3,2k-1}) + \\[3pt]
+ 2(\lambda_{2i+2}\wp_{2k}\delta_{i,1} + \lambda_{2k+2}\wp_{2i}\delta_{k,1}) + 2\lambda_{2(i+j+1)} (2\delta_{i,k} + \delta_{i,k-1} + \delta_{i-1,k}),
\end{multline}
were $\delta_{i,k}$ is the Kronecker symbol.
\end{theorem}

\begin{corollaryt}\label{C-76}
For any $k> 1$, the hyperelliptic Kleininan function $\wp_{1,2k-1}$ is a differential polynomial in $\wp_{2}$.
\end{corollaryt}

\begin{corollaryt}\label{C-77}
For any $g\geq 1$, we have
\begin{enumerate}
  \item [(a)] When $i=1$, Equation \eqref{f-55} implies $\wp''_{2} = 6\wp_{2}^2 + 4\wp_{4} + 2\lambda_4$.\\[5pt]
  \item [(b)] When $i=2$, Equation \eqref{f-55} implies $\wp''_{4} = 6(\wp_{2} \wp_{4} + \wp_{6}) - 2\wp_{3,3}$.\\[5pt]
  \item [(c)] When $i=k=1$, Equation \eqref{f-56} implies $(\wp'_{2})^2 = 4[\wp_{2}^3+(\wp_{4}+\lambda_4)\wp_{2}+\wp_{3,3}-\wp_{6}+\lambda_6]$.
\end{enumerate}
\end{corollaryt}

We have $\wp_{2i}' = \partial_{2i-1}\wp_{2}$. Then, from Corollary \ref{C-77}(a), we obtain
\begin{corollaryt}
For any $g>1$, the function $u = 2\wp_2(\mathbf{z})$ is a solution to the Korteweg--de Vries equation
\[
G''' = 6 G G' + 4\dot G,\; \text{ where }\; \dot G = \frac{\partial G}{\partial z_3}.
\]
\end{corollaryt}

\subsection{Solutions in Hyperelliptic Kleinian functions}\label{sect7.3} \text{}\\
The system of identities
\[
\partial_{2k-1}\wp_{2} = \partial_1\wp_{1,2k-1}
\]
defines the system of equations
\begin{equation}\label{eq-57}
\partial_{2k-1}u = \partial_1 B_{2k}(u),
\end{equation}
where $B_{2k}(u) = B_{2k}(u,u',\ldots,u^{(2k-2)}) = 2\wp_{1,2k-1}$ is a differential polynomial in $u$.

\begin{example}
\[
4\partial_3 \wp_{2} = \partial_1(\wp_{2}''-6\wp_{2}^2-2\lambda_4), \; \text{ and }\; B_4(u) = \frac{1}{4}(u''-3u^2-4\lambda_4).
\]
\end{example}

From statements Corollary \ref{C-77}(b--c), one can obtain an explicit form of the differential polynomial $B_6$.

\begin{theorem}\label{KdV-hkf-th}
System of equations \eqref{eq-57} coincides with the Korteweg--de Vries hierarchy for $\mathbf{t}=\mathbf{z}$ with parameters $\lambda_4,\ldots,\lambda_{2g+2}$ and integrals $\lambda_{2g+4},\ldots,\lambda_{4g-2}$,
which is solved by the hyperelliptic Klein function $\wp_{2}(\mathbf{z})$.
\end{theorem}

Let us put
\begin{align*}
p_\xi^{I} =&- \sum_{i=1}^{g} \wp_{2i}\xi^{g-i}+\xi^g ,  &u_\xi =&\; p_\xi^{I}, \\
p_\xi^{II} =&\;\;\;\, \sum_{i=1}^{g} \wp_{2i}'\xi^{g-i},  &v_\xi =&\; \frac{1}{2}p_\xi^{II}, \\
p_\xi^{III} =&\;\;\;\, \sum_{i=1}^{g} \wp_{2i}''\xi^{g-i},  &w_\xi =&\; (\xi+2\wp_{2})p_\xi^{II} + \frac{1}{2}p_\xi^{III}.
\end{align*}
From \cite[Theorem 8.4]{Buch23}, the matrix $L_\xi = 2
\begin{pmatrix}
  v_\xi & u_\xi \\
  w_\xi & -v_\xi
\end{pmatrix}$
satisfies the equation
\[
D_\eta L_\xi = \frac{1}{\xi-\eta}[L_\xi,L_\eta] + [L_\xi,A_\eta],
\]
where $A_\eta = -2
\begin{pmatrix}
  0 & 0 \\
  u_\eta & 0
\end{pmatrix}$. This equation coincides with the equation \eqref{f-1} of the Mumford system.

Using Theorem \ref{M-NS-th}, we obtain the solution to the Neumann--Moser system in hyperelliptic Kleinian functions.

\appendix

\vspace{1cm}

\section{Matrices in symmetric polynomials}\label{sectA}

Consider a vector $\phi = (\phi_1,\, \ldots,\, \phi_{n})$ for $n\geq 2$, where $\phi_i\neq \phi_j$ for all $1\leq i\neq j\leq n$.
\begin{definition}\label{symmetricdef1}
Define the \emph{elementary symmetric polynomials $e_k(\phi)$ of degree $k$}  for $1\leq k\leq n$  by
$$
e_1=\phi_1+\ldots+\phi_{n},\quad\ldots,\quad
e_k=\sum_{1\leq i_1<\ldots<i_k\leq n}\phi_{i_1}\cdots\phi_{i_k},\quad\ldots,\quad
e_n=\phi_{i_1}\cdots\phi_{i_n},$$
and the \emph{elementary symmetric $l$-polynomials $e^{(l)}_k(\phi)$ of degree $k$} for $1\leq l\leq n$ and $1\leq k\leq n-1$ by
$$
e^{(l)}_k=e_k(\phi^{(l)})=e_k(\phi_1,\,\ldots,\,\phi_{l-1},\,\phi_{l+1},\,\ldots,\,\phi_{n}).
$$
\end{definition}

\begin{definition}\label{symmetricdef1-1}
Define the \emph{complete homogeneous symmetric polynomials $E_k(\phi)$ of degree $k$} for $1\leq k\leq n$  by
$$E_k=\sum_{1\leq i_1\leq\ldots\leq i_k\leq n}\phi_{i_1}\cdots\phi_{i_k},$$
and the \emph{complete homogeneous symmetric $l$-polynomials $E^{(l)}_k(\phi)$ of degree $k$} for $1\leq l\leq n$ and $1\leq k\leq n-1$ by
$$
E^{(l)}_k=E_k(\phi^{(l)})=E_k(\phi_1,\,\ldots,\,\phi_{l-1},\,\phi_{l+1},\,\ldots,\,\phi_{n}).
$$
\end{definition}

Consider $(n+1)\times(n+1)$-matrices
\begin{equation*}
D(\phi)=
\begin{pmatrix}
  1 & 0 & \ldots & 0 & 0\\
  e_1(\phi) & 1 & \ldots & 0 & 0\\
  e_2(\phi) & e_1(\phi) & \ldots & 0 & 0\\
 \vdots & \vdots & \ddots & \vdots & \vdots \\
  e_{n-1}(\phi) & e_{n-2}(\phi) & \ldots & 1 & 0\\
  e_{n}(\phi)   & e_{n-1}(\phi) & \ldots & e_1(\phi) & 1
\end{pmatrix},
\end{equation*}
\begin{equation*}
\overline{\!D}(\phi)=
\begin{pmatrix}
  1 & 0 & \ldots & 0 & 0 & 0\\
  -E_1(\phi) & 1 & \ldots & 0 & 0 & 0\\
  E_2(\phi) & E_1(\phi) & \ldots & 0 & 0 & 0\\
 \vdots & \vdots & \ddots & \vdots & \vdots & \vdots \\
  (-1)^{n-1}E_{n-1}(\phi) & (-1)^{n-2}E_{n-2}(\phi) & \ldots & -E_1(\phi) & 1 & 0\\
  (-1)^{n}E_{n}(\phi)   & (-1)^{n-1}E_{n-1}(\phi) & \ldots & E_2(\phi) & -E_1(\phi) & 1
\end{pmatrix}.
\end{equation*}

\begin{lemma}\label{inverse_D_lemma}
$$
\overline{\!D}(\phi) = D^{-1}(\phi).
$$
\end{lemma}
\begin{proof}
  Consider $F=D(\phi)\cdot \overline{\!D}(\phi)$. We want to calculate all the matrix elements $f_{i,j}$ for $1\leq i$,\, $j\leq n+1$. There are 5 cases to consider. Obviously, $f_{i,i}=1$ for all $1\leq i\leq n+1$ and $f_{i,j}=0$ for all $1\leq i<j\leq n+1$. It remains to consider the case of $1\leq j<i\leq n+1$.
\begin{multline*}
f_{i,j}=e_{i-j}(\phi)\cdot 1+\sum_{k=j+1}^{i-1}e_{i-k}(\phi)\cdot(-1)^{k-j}E_{k-j}(\phi) +1\cdot (-1)^{i-j}E_{i-j}(\phi)=\\
=e_{i-j}(\phi)+\sum_{l=1}^{i-j-1}(-1)^{l}E_{l}(\phi)\cdot e_{i-j-l}(\phi) +(-1)^{i-j}E_{i-j}(\phi)=0.
\end{multline*}
\end{proof}


\begin{definition}\label{symmetricdef2}
Define the \emph{Vandermonde polynomial $\Pi(\phi)$} and the
\emph{Vandermonde $l$-polynomials $\Pi^{(l)}(\phi)$}, $1\leq l\leq n,$ by
$$
\Pi(\phi)=\prod_{1\leq i<j \leq n}(a_j-a_i),\quad
\Pi^{(l)}(\phi)=\Pi(\phi^{(l)})=\prod_{\substack{1\leq i<j \leq n\\ i,\,j\neq l}}(a_j-a_i).
$$
\end{definition}

Consider $(n+1)\times(n+1)$-matrices
$$
M(\phi)=
\left(\begin{array}{ccccc}
1 & 0 & 0 & \ldots & 0 \\
e_1(\phi) & 1 & 1 & \ldots & 1 \\
e_2(\phi) & e_1^{(1)}(\phi) & e_1^{(2)}(\phi) & \cdots & e_1^{(n)}(\phi) \\
\vdots & \vdots & \vdots & \ddots & \vdots \\
e_{n}(\phi) & e_{n-1}^{(1)}(\phi) & e_{n-1}^{(2)}(\phi) & \cdots & e_{n-1}^{(n)}(\phi)
\end{array}\right).
$$
and
$$
\overline{\!M}(\phi)=\frac{1}{\Pi(\phi)}
\left(\begin{array}{cccc}
\Pi(\phi) & \ldots & 0 & 0 \\
(-1)^1\phi_1^{n}\Pi^{(1)}(\phi) & \ldots & (-1)^{n}\phi_1\Pi^{(1)}(\phi) & (-1)^{n+1}\Pi^{(1)}(\phi) \\
(-1)^2\phi_2^{n}\Pi^{(2)}(\phi)  & \ldots & (-1)^{n+1}\phi_2\Pi^{(2)}(\phi) & (-1)^{n+2}\Pi^{(2)}(\phi) \\
\vdots   & \ddots & \vdots & \vdots \\
(-1)^{n}\phi_n^{n}\Pi^{(n)}(\phi) & \ldots & (-1)^{2n}\phi_n\Pi^{(2n)}(\phi) & (-1)^{2n+1}\Pi^{(n)}(\phi)
\end{array}\right).
$$
For $2\leq i\leq n+1$, the $(i,j)$-entry of $\overline{\!M}(\phi)$ is $\bar{m}_{ij}=(-1)^{j+i-1}\phi_{i-1}\Pi^{(i-1)}(\phi)$.

\begin{lemma}\label{inverse_M_lemma}
$$
\overline{\!M}(\phi) = M^{-1}(\phi).
$$
\end{lemma}
\begin{proof} Consider $N(\phi)=\overline{\!M}(\phi)\cdot M(\phi)$. We want to calculate all the matrix elements $n_{i,j}$ for $1\leq i$,\, $j\leq n+1$. There are 5 cases to consider.
\begin{enumerate}
\item $i=1$, $j=1$:
$$
n_{1,1}=\frac{1}{\Pi(\phi)}\Big[\Pi(\phi)\cdot 1 + \sum_{k=1}^{n}0\cdot e_k(\phi)\Big]=1.
$$

\item $i=1$, $j\neq 1$:
$$
n_{1,j}=\frac{1}{\Pi(\phi)}\Big[\Pi(\phi)\cdot 0 + 0\cdot 1 + \sum_{k=1}^{n-1}0\cdot e_k^{(j)}(\phi)\Big]=0.
$$

\item $i\neq 1$, $j=1$:
\begin{multline*}
n_{i,1}=\frac{(-1)^{i}}{\Pi(\phi)}\Big[-\phi_i^{n}\Pi^{(i)}(\phi)\cdot 1 + \sum_{k=1}^{n}(-1)^{k+1}\phi_i^{n-k}\Pi^{(i)}(\phi)\cdot e_k(\phi)\Big]= \\
=\frac{(-1)^i}{\Pi(\phi)}\Big( \phi_i^{n} + \sum_{k=1}^{n}(-1)^{k+1}\phi_i^{n-k}\cdot e_k(\phi) \Big)=0.
\end{multline*}

\item $1\neq i\neq j\neq 1$:
\begin{multline*}
n_{i,j}=\frac{(-1)^{i}}{\Pi(\phi)}\Big[-\phi_i^{n}\Pi^{(i)}(\phi)\cdot 0 + \phi_i^{n-1}\Pi^{(i)}(\phi)\cdot 1 +  \sum_{k=1}^{n-1}(-1)^{k}\phi_i^{n-1-k}\Pi^{(i)}(\phi)\cdot e^{(j)}_k(\phi)\Big]=\\
=\frac{\Pi^{(i)}(\phi)}{\Pi(\phi)}\Big[ \phi_i^{n-1} + \sum_{k=1}^{n-1}(-1)^{k}\phi_i^{n-1-k}e^{(j)}_k(\phi) \Big]=0.
\end{multline*}

\item $1\neq i=j\neq 1$:
\begin{multline*}
n_{i,i}=\frac{(-1)^{i}}{\Pi(\phi)}\Big[-\phi_i^{n}\Pi^{(i)}(\phi)\cdot 0 + \phi_i^{n-1}\Pi^{(i)}(\phi)\cdot 1 + \sum_{k=1}^{n-1}(-1)^{k}\phi_i^{n-1-k}\Pi^{(i)}(\phi)\cdot e^{(i)}_k(\phi)\Big]=\\
=\frac{(-1)^{i}\Pi^{(i)}(\phi)}{\Pi(\phi)}\Big[ \phi_i^{n-1} + \sum_{k=1}^{n-1}(-1)^{k}\phi_i^{n-1-k}e^{(i)}_k(\phi) \Big]=\\
=\frac{(-1)^{i}\Pi^{(i)}(\phi)}{\Pi(\phi)}\Big[ \phi_i^{n-1} - e^{(i)}_1(\phi)\phi_i^{n-2} +\ldots + (-1)^{n-2}e^{(i)}_{n-1}(\phi)\phi_i+ (-1)^{n-1}e^{(i)}_{n}\Big].$$
\end{multline*}
\end{enumerate}
Consider the polynomial
$$
P(\phi_i)=\phi_i^{n-1} - e^{(i)}_1(\phi)\phi_i^{n-2} +\ldots + (-1)^{n-2}e^{(i)}_{n-1}(\phi)\phi_i+ (-1)^{n-1}e^{(i)}_{n}.
$$
Note that $e^{(i)}_1(\phi)$ does not depend on $\phi_i$. Therefore, this is a degree $(n-1)$ polynomial in $\phi_i$. Moreover, it is easy to see that
$$
P(\phi_i)=\prod_{1\leq j\leq n}^{j\neq i} (\phi_i-\phi_j).
$$
Therefore, $n_{i,i}=1$.
\end{proof}

\begin{corollaryt}
$\det M(\phi) =\Pi(\phi)$.
\end{corollaryt}

\vspace{1cm}



\end{document}